\documentclass[amssymb,amsmath,aps,showpacs,floatfix,nofootinbib,showpacs,12pt]{revtex4}
\usepackage{graphicx,amssymb,color,soul,latexsym}

\begin{document}

\title{Discriminating different scenarios to account for the cosmic $e^\pm$ excess
 by synchrotron and inverse Compton radiation }

\author{Juan Zhang$^{1}$, Xiao-Jun Bi$^{1,2}$, Jia Liu$^{3}$, Si-Ming Liu$^{4}$, Peng-Fei Yin$^{3}$, Qiang Yuan$^{1}$ and Shou-Hua Zhu$^{3}$}
\affiliation{$^{1}$ Key Laboratory of
Particle Astrophysics, Institute of High Energy Physics, Chinese
Academy of Sciences, Beijing 100049, P. R. China \\
$^{2}$ Center for High Energy Physics,
Peking University, Beijing 100871, P.R. China \\
$^{3}$ Institute of Theoretical Physics \& State Key
Laboratory of Nuclear Physics and Technology, Peking University,
Beijing 100871, P.R. China \\
$^{4}$ Department of Physics and Astronomy, University of Glasgow, \\
Kelvin Bldg Rm 620, Glasgow G12 8QQ, United Kingdom
}

\date{\today}

\begin{abstract}

The excesses of the cosmic positron fraction recently measured by
PAMELA and the electron spectra by ATIC, PPB-BETS, Fermi and
H.E.S.S. indicate the existence of primary electron and positron
sources. The possible explanations include dark matter
annihilation, decay, and astrophysical origin, like pulsars. In
this work we show that these three scenarios can all explain the
experimental results of the cosmic $e^\pm$ excess. However, it may
be difficult to discriminate these different scenarios by the
local measurements of electrons and positrons. We propose possible
discriminations among these scenarios through the synchrotron and
inverse Compton radiation of the primary electrons/positrons from
the region close to the Galactic center. Taking
typical configurations, we find the three scenarios predict quite
different spectra and skymaps of the synchrotron and inverse
Compton radiation, though there are relatively large
uncertainties. The most prominent differences come from the energy
band $10^4\sim 10^9$ MHz for synchrotron emission and $\gtrsim 10$
GeV for inverse Compton emission. It might be able to discriminate
at least the annihilating dark matter scenario from the other two
given the high precision synchrotron and diffuse $\gamma$-ray
skymaps in the future.

\end{abstract}



\maketitle

\section{Introduction}

The anti-matter particles such as positrons and antiprotons in
cosmic rays (CRs) are very important in understanding the origin
and propagation of CRs. These particles are usually produced by
the CR nuclei that interact with the interstellar medium (ISM)
when propagating in the interstellar space. Precise measurements
of these particles will provide us valuable information about the
primary CR sources and the interaction with matter. In addition,
these secondary anti-matter particles have relative lower fluxes
and characteristic spectra. Therefore they make themselves good objects
to study the exotic origin of CRs, such as from dark
matter (DM) annihilation or decay.

Recently the PAMELA collaboration released their first CR
measurements on the positron fraction \cite{Adriani:2008zr} and
$\bar{p}/p$ ratio \cite{Adriani:2008zq}. The positron fraction of
PAMELA data shows an uprise above $\sim 10$ GeV till to $\sim 100$
GeV, which exceeds the background estimation of the conventional
CR propagation model. This result is consistent with previous
measurements by, e.g., HEAT \cite{Barwick:1997ig} and AMS
\cite{Aguilar:2007yf}. On the other hand, the $\bar{p}/p$ ratio is
compatible with the background. The results of PAMELA strongly
indicate the existence of such sources that mainly generate
leptonic particles at this energy range. The electron spectrum up
to several TeV reported by ATIC collaboration also shows an
obvious excess with an interesting bump around $300\sim800$ GeV
\cite{Chang:2008zz}. In addition, the measurements of
the electron spectrum by PPB-BETS \cite{Torii:2008xu}, H.E.S.S.
\cite{Collaboration:2008aa,Aharonian:2009ah}, and most recently by
Fermi \cite{Collaboration:2009zk} all show the excesses of
electron spectra, although they are not fully consistent between
each other. The studies show that the PAMELA result on the
positron fraction is consistent with the electron spectrum
measured by ATIC or Fermi assuming the equal amount of the
production of positrons and electrons (e.g.,
\cite{Cholis:2008wq,Bergstrom:2009fa}).

One possible primary electron/positron source in the Galaxy is
the pulsar. Pulsars and their nebulae are well known cosmic
particle accelerators. Although the quantitative details of the
acceleration processes are still open for study, early radio
observations have established them as important high energy cosmic
electron and positron sources (e.g., \cite{Frail1997}). X-ray and
$\gamma$-ray observations show that some of the accelerated
particles can reach an energy of a few tens of TeV, and there are
indications that the particle distribution cuts off in the TeV
energy range (e.g., \cite{Pavlov2001,Lu:2002kx,
Aharonian:2006xx}). These particles produce emission over a
broadband frequency range in the source region, which has been
observed and studied extensively (e.g., \cite{Li:2007dg}). Some of
these particles will escape into the ISM becoming high energy
cosmic electrons and positrons. It is shown that one or several
nearby pulsars will be able to contribute enough positrons to
reproduce the PAMELA data \cite{Hooper:2008kg,Yuksel:2008rf}
as well as ATIC data \cite{Profumo:2008ms}.

Another potential candidate, which has been widely discussed to
solve the positron/electron excess problem, is the DM annihilation (e.g.,
\cite{annidm}) or decay (e.g.,\cite{decaydm,Yin:2008bs,Nardi:2008ix}).

It has been shown in many studies that the local
measurements of the electrons/positrons can not provide enough
power to discriminate the DM origin and the pulsar origin of the
electrons/positrons. On one hand, the ATIC data suggest a sharp
cutoff near $\sim 600$ GeV. Such a feature can naturally appear
for a DM origin when closing to the mass of DM particle, if the DM
annihilates or decays directly to e$^+$e$^-$ pair, as shown in
Ref. \cite{Chang:2008zz}. However, the same thing is also expected
if the observed excess is dominated by a single pulsar and/or its
nebula and the accelerated high energy electrons and positrons
have experienced significant energy loss as they propagate in the
source and in the ISM \cite{Profumo:2008ms}. The narrow peak
resulting from local pulsars is even indistinguishable from that
of DM origin \cite{Pohl:2008gm}. On the other hand, if the
electron spectrum is as smooth as that from Fermi, there are also
degeneracies between the pulsar and DM interpretations. A
continuously distributed pulsar population in the Galaxy might be
able to generate smoother electron spectrum due to different
energy losses of various pulsars. While for the DM scenario, if
the electrons and positrons come from DM annihilation or decay,
the spectrum can also be smooth, and is consistent with the Fermi
measurements \cite{Bergstrom:2009fa}.

In this work we study the photon emission associated with the
different scenarios to see how the degeneracies between these
models can be broken. Our point is that although the different
scenarios can give almost the same local electrons/positrons, the
spatial extrapolations of them might be significantly different,
especially in the Galactic center (GC) region. The pulsars mainly
concentrate in the Galactic plane, while the DM is spherically
distributed in an extended halo and highly concentrated in the
center of the halo. In addition, DM annihilation is proportional
to $\rho^2(r)$, different from that of DM decay ($\propto
\rho(r)$). Therefore we anticipate the existence of remarkable
differences in the skymaps of the synchrotron and inverse Compton
(IC) emission from these extra primary electrons/positrons for the
three scenarios. 

We compare the differences of synchrotron as well as IC radiation
among the pulsars, annihilating DM and decaying DM models, on the
premise that they all reproduce the positron fraction and total
electron + positron flux data. The spectral and directional
features of the radiation are discussed in detail. We compare the
results with the WMAP haze data
\cite{Finkbeiner:2003im,Dobler:2007wv} and EGRET diffuse
$\gamma$-ray data \cite{Hunter:1997we}, and give the observable
signals by future experiments such as Fermi/GLAST.

This paper is organized as follows. In Sec. II we will briefly
present typical model configurations of the three scenarios which
can fit the $e^\pm$ data simultaneously. The discussions
on the synchrotron and IC emission are given in Sec. III and IV
respectively. In Sec. V we discuss the uncertainties of our
predictions. Finally, we give a summary and draw the conclusion in
Sec. VI.

\section{Different scenarios to account for the $e^\pm$ data}


We start with the propagation equation of electrons/positrons in the
Galaxy
\begin{eqnarray}
\frac{\partial \psi}{\partial t} =Q({\bf
r},p)+\nabla\cdot(D_{xx}\nabla \psi-{\bf
V_c}\psi)+\frac{\partial}{\partial p}p^2D_{pp}\frac{\partial}
{\partial p}\frac{1}{p^2}\psi \nonumber \\
 - \frac{\partial}{\partial p}\left[\dot{p}\psi
-\frac{p}{3}(\nabla\cdot{\bf V_c}\psi)\right]
\ , \label{prop}
\end{eqnarray}
where $\psi$ is the number density of CR particles per unit
momentum interval, $Q({\bf r},p)$ is the source term, describing
the primary particles injected into the interstellar medium,
$D_{xx}$ is the spatial diffusion coefficient, ${\bf V_c}$ is the
convection velocity. The reacceleration process is described by
the diffusion in momentum space, with the diffusion coefficient
$D_{pp}=\frac{4p^2v_A^2}{3\delta(4-\delta^2)(4-\delta)wD_{xx}}$,
where $v_A$ is the Alfven speed, $w$ is the ratio of
magnetohydrodynamic wave energy density to the magnetic field
energy density, which characterizes the level of turbulence.
$\dot{p}\equiv{\rm d}p/{\rm d}t$ is the momentum loss rate, mainly
induced by synchrotron radiation and IC scattering for the
electrons/positrons for energies $\gtrsim 10$ GeV
\cite{Strong:1998pw}. In this work, we use the numerical package
GALPROP \cite{Strong:1998pw} to calculate the propagation of the
primary electrons/positrons and the background CRs.

We have adopted the conventional propagation model where all the
CR data are reproduced by the model. The cosmic nuclei secondary
to primary ratio, such as B/C, the unstable secondary to stable
secondary, such as $^{10}$Be/$^9$Be, and the local proton and
electron spectra are taken to constrain the propagation
parameters. A conventional diffusion + convection (DC) propagation
model \cite{Yin:2008bs} is adopted. For details of the propagation
model see the discussion in Ref. \cite{Yin:2008bs}. The
propagation parameters are taken as: half height of the
propagation halo $z_h=4$ kpc; diffusion coefficient $D_{xx}=\beta
D_0 (\rho/\rho_0)^\delta$ with
$D_0=2.5\times10^{28}\,$cm$^2\,$s$^{-1}$, $\delta=0.55$ for
rigidity $\rho>4$ GV and $\delta=0$ for $\rho<4$ GV; the
convection velocity is adopted as a linear function of coordinate
$z$ with $V_c(z=0)=0$ and ${\rm d}V_c/{\rm d}z=6$ km s$^{-1}$
kpc$^{-1}$. The reacceleration is not included.

The source term $Q({\bf r},p)$ is different for various scenarios.
For DM annihilation, the source function of electrons/positrons has
the form
\begin{equation}
\label{anni}
Q_{A}({\bf r},E)={\rm BF}\frac{\langle\sigma v\rangle_{A} \rho^2(r)}
{2\,m^2_{DM}}\left.\frac{dN(E)}{dE}\right|_{A},
\end{equation}
where BF represents the ``boost factor'' which can come from the
clumpiness of DM substructures (e.g., \cite{Yuan:2006ju,Bi:2007ab,
Bi:2006vk}) and/or the so-called Sommerfeld effect \cite{annidm}.
In the former case, BF is energy and spatial dependent, ${\rm BF}\sim
{\rm BF}({\bf r},E)$. However, the study based on
N-body simulations shows that the DM substructures do not tend to
strongly enhance the DM annihilation signals \cite{Lavalle:1900wn}. 
Thus in this work we adopt the Sommerfeld effect as the ``boost factor''
for discussion. $\langle\sigma v\rangle_{A}$ is the velocity weighted
annihilation cross section before being enhanced by the Sommerfeld effect,
$m_{DM}$ is the mass of DM particle, $\frac{dN(E)}{dE}\mid_{A}$ is the
electron/positron spectrum from one pair of DM annihilation, and $\rho(r)$
is DM density distribution in the Galaxy. To satisfy the relic density
constraint we adopt the cross section $\langle\sigma v\rangle_{A}=3\times
10^{26}$ cm$^3$ s$^{-1}$. For the DM density profile we adopt the Merritt
profile \cite{Merritt:2005xc}
\begin{equation}
\rho(r)=\rho_0\exp\left[-\frac{2}{\alpha}\left(\frac{r^{\alpha}-
R^{\alpha}_{\odot}}{r^{\alpha}_{-2}}\right)\right],
\end{equation}
with $\alpha=0.2$, $r_{-2}=25$ kpc and the local DM density
$\rho_{\odot}=0.3$ GeV cm$^{-3}$ \cite{Cholis:2008vb}
at $r=R_{\odot}\equiv8.5\,$kpc.

For DM decay, the source term is given by
\begin{equation}
Q_{D}({\bf r},E)=
\frac{1}{\tau_{DM}}\frac{\rho(r)}{m_{DM}}\left.\frac{dN}{dE}\right|_{D}
\label{sourcedecay}
\end{equation}
where $\tau_{DM}$ is the life time of DM, $\frac{dN}{dE}\mid_{D}$ is
the electron/positron spectrum from the decay of one DM particle.

The source term for Galactic pulsars is
\begin{equation}
 Q_P (R,z,E) = K \cdot f(R,z) \cdot
\left.\frac{{dN}}{{dE}}\right|_P\ , \label{sourcepulsar}
\end{equation}
where $K$ is the normalization factor representing the total luminosity
of Galactic pulsars, $f(R,z)$ is the pulsar spatial distribution and
$\frac{{dN}}{{dE}}|_P$ is the average electron/positron energy
spectrum radiated from the pulsars. The spatial distribution of
pulsars can be parameterized as
\begin{equation}
f(R,z)\propto\left(\frac{R}{R_{\odot}}\right)^{a}\,
\exp\left[{-\frac{b(R-R_{\odot})}{R_{\odot}}}\right]
\exp\left({-\frac{|z|}{z_s}}\right),
\end{equation}
where $R$ is the  Galactocentric radius, and $z$ is the distance
away from the Galactic Plane. Different from spherically symmetric form
of DM distribution, pulsars are mainly concentrated at the Galactic Plane
with $z_s\sim 0.2$ kpc.
The primary electron/positron spectrum injected by pulsars is generally
assumed to be a power law form with an exponential cutoff at high energies
\begin{equation}
\left.\frac{{dN}}{{dE}}\right|_P = E^{ - \alpha }\exp\left(-E/E_{cut}\right) ,
\end{equation}


\begin{figure}[!htb]
\centering
\includegraphics[width=0.45\columnwidth]{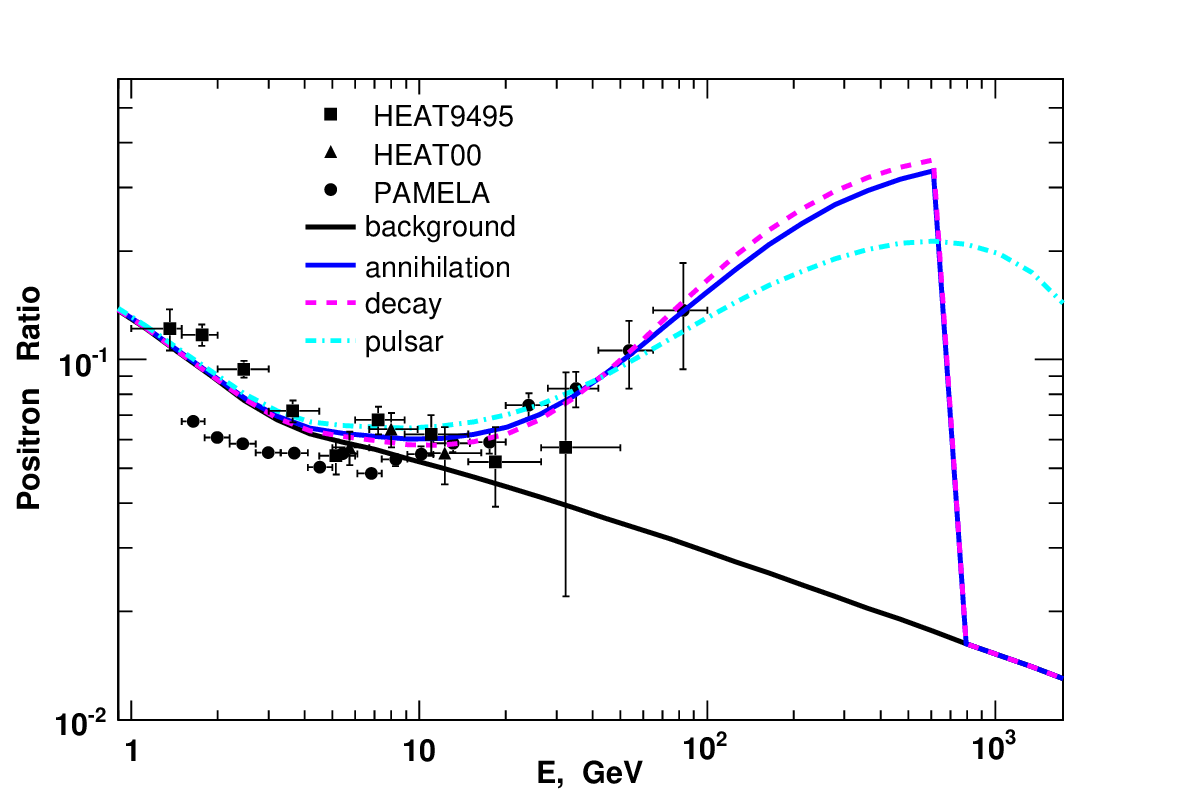} 
\includegraphics[width=0.45\columnwidth]{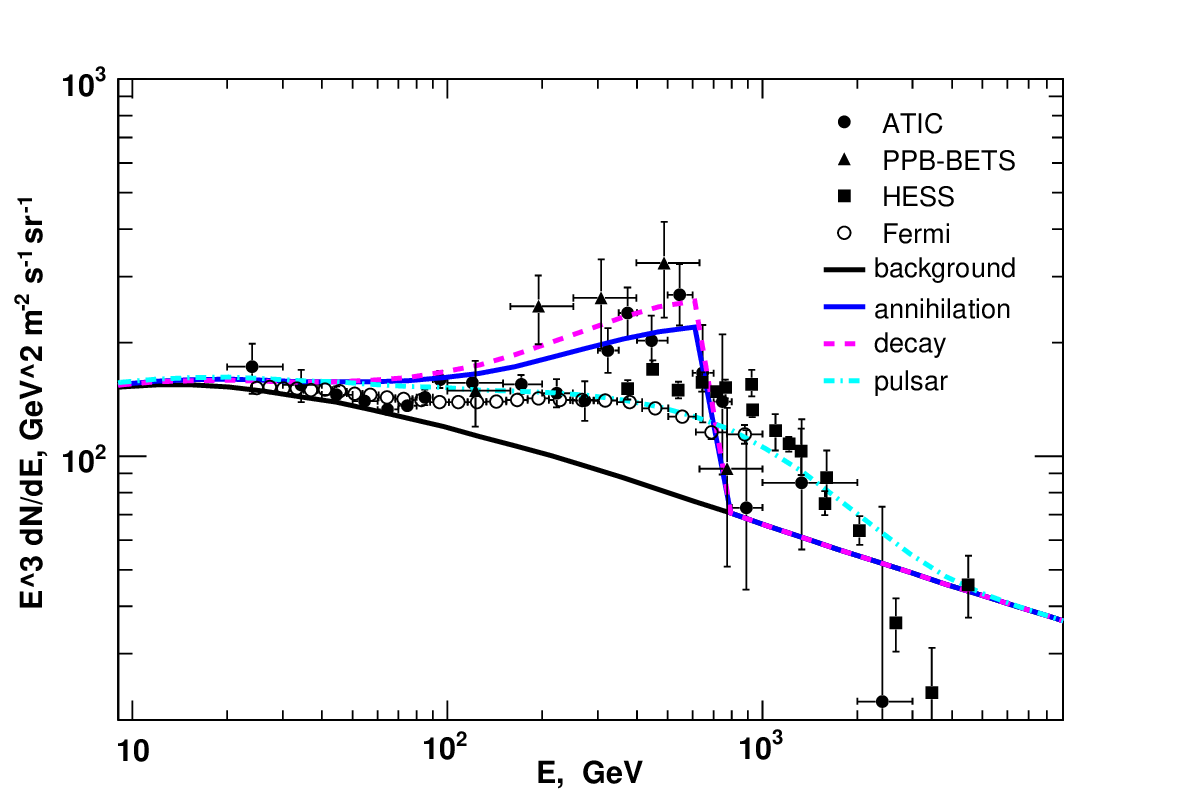}
\caption{ {\it Left:} The positron fractions predicted by the
three scenarios after solar modulation, compared with  HEAT
\cite{Barwick:1997ig,Coutu:2001jy} and PAMELA
\cite{Adriani:2008zr} data. {\it Right:} The total 
electron+positron fluxes  
of the three scenarios, compared with observations of ATIC
\cite{Chang:2008zz}, PPB-BETS \cite{Torii:2008xu}, H.E.S.S.
\cite{Collaboration:2008aa,Aharonian:2009ah} and Fermi
\cite{Collaboration:2009zk}. We assume DM annihilates or decays
into pure lepton final states with equal branching ratios to
$e^+e^-$, $\mu^+\mu^-$ and $\tau^+\tau^-$.
}.\label{positrons}
\end{figure}

We have taken appropriate parameters in each scenario to give best fit to
the PAMELA and ATIC or Fermi data. For the annihilating DM, we adopt
$m_{DM}=1$ TeV, and a boost factor $BF = 800$. The annihilation
channels are assumed to be pure leptons with equal branching
ratios to $e^+e^-$, $\mu^+\mu^-$ and $\tau^+\tau^-$ respectively.
We find that the PAMELA and ATIC data can not be fitted well by
only one lepton flavor final state. Note, however, for the
case of Fermi data the pure $\mu^+\mu^-$ or $\tau^+\tau^-$ channel might
be able to fit the data. We will further discuss this point in Sec. V. 
For the decaying DM, we set DM mass $m_{DM}=2\,$TeV and life time $\tau
\sim 1.08\times 10^{26}$s. The decay channels are the same as the
annihilation scenario. For pulsars, we adopt the spatial distribution
parameters $a=1.0$, $b=1.8$ \cite{zhangli}, and the power law index of
the energy spectrum $\alpha \sim 1.4$ with an exponential cutoff above
$\sim 800$ GeV. The spectrum index $1.4$ can give good description to
the Fermi data. However, to better fit the ATIC data we need to introduce
harder spectrum and/or super-exponential cutoff.
%

The positron fraction and the total electron + positron 
flux are shown in Figure \ref{positrons}.
The solar modulation effect is calculated using the force field
approximation \cite{gleeson68} with a potential of 500\,MV in this work.
For the background electron choice, we adopt the DC model given in Ref.[22]. 
A rescale factor ~0.9 on the electron normalization is adopted to better fit 
the PAMELA and ATIC/Fermi data simultaneously. This model is actually similar 
with the so-called conventional model \cite{Strong:2004de},  
which fits all the pre-Fermi CR data and 
is also successfully used to model the diffuse gamma-ray spectrum 
measured by Fermi at intermediate Galactic latitudes \cite{Grasso:2009ma}. 

Finally we should point out that we take
continuous distribution of the sources (pulsar or DM) to recover
the PAMELA and ATIC/Fermi data in this work. Note that it is also
possible that the locally observed electrons/positrons come from
one or several nearby pulsars \cite{Hooper:2008kg,Profumo:2008ms}
or DM clumps \cite{Hooper:2008kv}. For the pulsar scenario, this
assumption does not affect the discussion since the extrapolation
to the whole Galaxy will hold regardless of how much the far
pulsars can contribute to the local electron flux. For the
annihilating DM model, although a single DM clump which is close
enough to the Earth may be possible to give large boost factor, it
is found to be of little probability to survive in a realistic DM
distribution model \cite{Lavalle:2006vb,Brun:2009aj}. For the
decaying DM scenario, since the ``boost'' effect from DM decay
will be much weaker we will expect an even smaller probability to
find a clump close and massive enough to explain the
electron/electron data, compared with the annihilating DM
scenario. Therefore the assumption of continuous distribution of
sources is reasonable.

\section{Synchrotron radiation from the three scenarios}



In this section we study the synchrotron emission from the primary
electrons and positrons in the three scenarios to account for the
locally observed cosmic $e^\pm$ excesses. The predictions are
based on the same configurations as discussed in the last section.

The synchrotron emissivity with frequency $\nu$ of a single
electron with Lorentz factor $\gamma$ in a magnetic field
$B$, is given as \cite{Ghisellini:1988}
\begin{equation}
\epsilon(\nu,\gamma)=4\sqrt{3}\pi r_e m_e c \nu_L x^2
\left\{K_{4/3}(x)K_{1/3}(x)-\frac{3x}{5}\left[K^2_{4/3}(x)-
K^2_{1/3}(x)\right]\right\} {\,\,\rm ergs\ s}^{-1}{\rm Hz}^{-1}\,
\label{emissivity}
\end{equation}
where $r_e=e^2/m_e c^2$ is the classical electron radius,
$x\equiv \nu/(3 \gamma^2 \nu_L)$ with $\nu_L=eB/2\pi m_e c$ the Lamor
frequency, and $K_l(x)$ is the modified Bessel function of order $l$.
For an ensemble of electrons/positrons with power law spectrum
$\frac{{dN}}{{dE}}\propto E^{-\alpha}$, the corresponding synchrotron
intensity is also a power law form $I(\nu)\propto\nu^{-\beta}$,
with index $\beta=(\alpha-1)/2$. The magnetic field strength in the
Galaxy is adopted as \cite{Strong:1998fr}
\begin{equation}
B(R,z)=B_0\exp\left(-\frac{R-R_\odot}{R_B}\right)
\exp\left(-\frac{|z|}{z_B}\right),
\end{equation}
where $R_B=10$ kpc and $z_B=2$ kpc, and the local magnetic field strength
$B_0=5\,\mu$G. Such a magnetic field can match the 408 MHz synchrotron
longitude and latitude distributions \cite{Strong:1998fr}.

Solving Eq. (\ref{prop}) we get the electron/positron
distributions in the Galaxy. Then we can derive the synchrotron
emissivity at any point in the Galaxy. The line of sight integral
of the emissivity gives the synchrotron flux as a function of
direction $(l,b)$, which can be compared with the observation.

\subsection{Results from the three scenarios}

\begin{figure}[!htb]
\centering
\includegraphics[width=0.45\columnwidth]{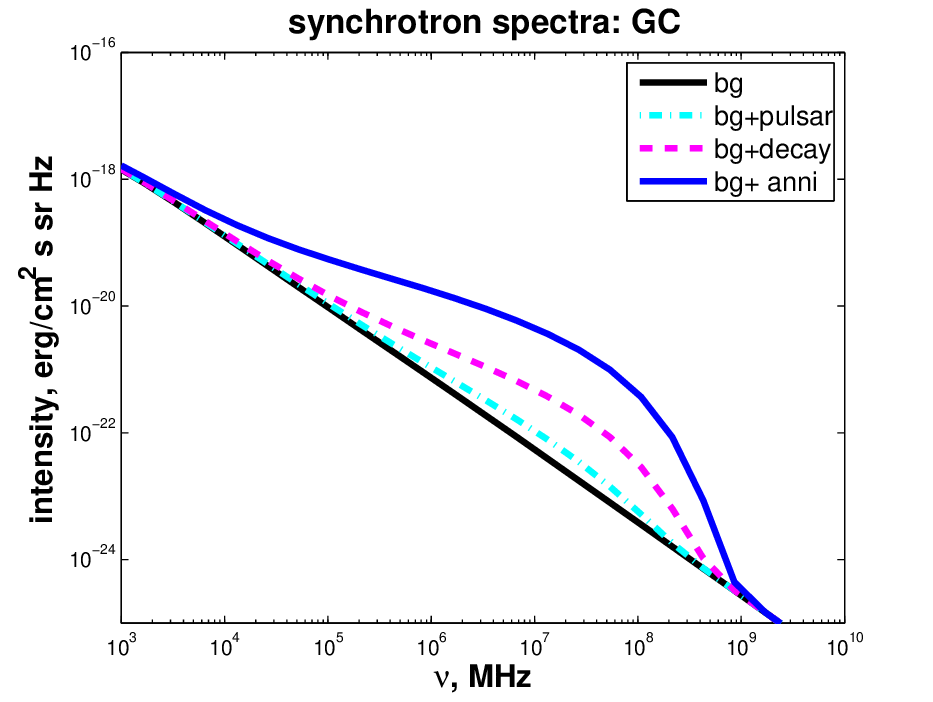} 
\includegraphics[width=0.45\columnwidth]{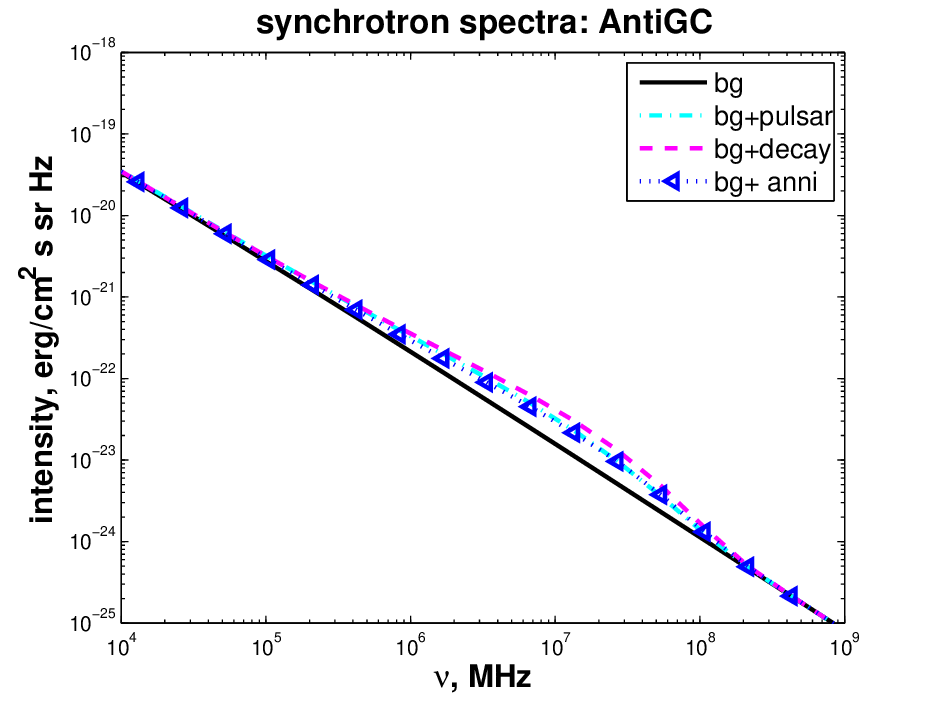} 
 \caption{The average synchrotron spectra of the three scenarios within
a bin size of $20^\circ\times20^\circ$ around GC and anti-GC, respectively.
} \label{spectra}
\end{figure}

In Figure \ref{spectra}, we give the average synchrotron spectra
of the three different scenarios in a $20^\circ\times20^\circ$
window centered at the Galactic center (GC) and anti-GC
respectively. In the GC direction the annihilating DM scenario has
the largest signal, while in the anti-GC direction the decaying DM
model has the largest signal though not very high in absolute
fluxes. It is shown that these models have quite different
synchrotron radiation spectra in the GC direction from $10^4$ MHz
to $10^9$ MHz, even though they have similar contributions to the
local electrons/positrons.

We give the results by smoothing the signals within the
$20^\circ\times20^\circ$ window centered at the GC so that we can
diminish the uncertainties of our prediction. The largest
uncertainty comes from the DM density profiles. The prediction of
DM annihilation at the GC from cuspy or cored density profile may
be orders of difference. By smoothing the signal within a large
window the uncertainty is considerably decreased (see detailed
discussion in Sec. 5). Further with a large window we can treat
the pulsars as continuous distribution with average emission
spectrum. Uncertainties from background are also deceased in a
large window.

\begin{figure}[!htb]
\centering
\includegraphics[width=0.45\columnwidth]{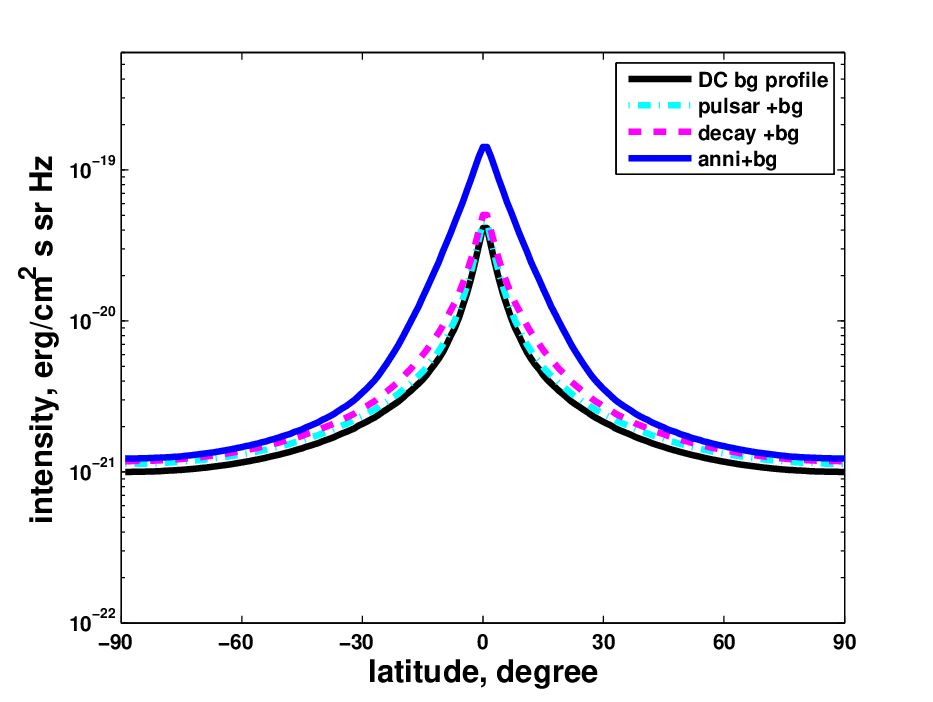}
\includegraphics[width=0.45\columnwidth]{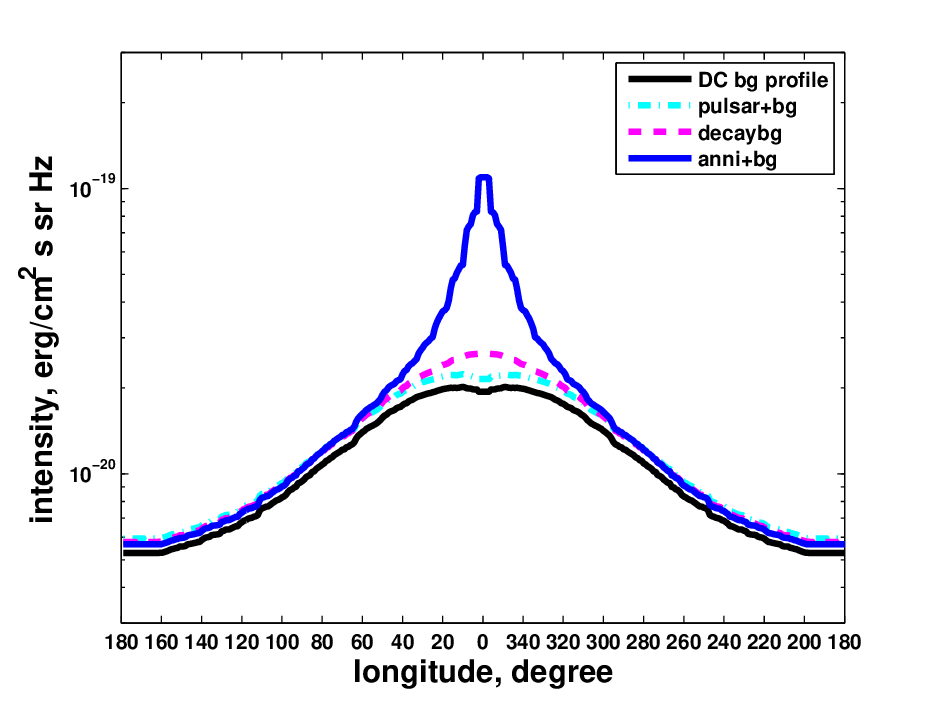}
\caption{
The synchrotron latitude profile for $|l|<10^\circ$ (left) 
and longitude profile for $|b|<10^\circ$ (right), at the 
frequency of 61 GHz. The solid black line indicates the 
contribution from background electrons and positrons, and 
the other lines represent the total synchrotron intensities 
from the three scenarios together with the background respectively.
} \label{pf}
\end{figure}

Given the large difference of synchrotron radiation
from $10^4$ MHz to $10^9$ MHz in Figure \ref{spectra},
we choose a frequency randomly in this range, $61$ GHz, to demonstrate
the synchrotron longitude and latitude profiles in Figure \ref{pf}.
It is
shown that the DM model will have larger gradient around GC, especially for
the longitude profile, when compared with the pulsar model. For
annihilating DM model the longitude profile is extremely cuspy
around $l=0$. This is a significant feature to distinguish among
these models. In addition, the absolute fluxes of the synchrotron
radiation in the inner Galaxy are also different. For pulsar
model, the longitude and latitude profiles almost follow the
background distributions and might be difficult to be distinguished
from the background. It is not hard to understand these profile
features. The pulsar distribution is similar to the source
distribution of CRs such as the supernova remnants, while the DM
distribution is spherically symmetric and much more concentrated
at GC than the pulsars. In addition the DM annihilation products
are proportional to the square of DM density, $\rho^2(r)$, while
the decay products depend only on the DM density $\rho(r)$. If the
DM density is cuspy in the center of the halo, like that shown by
many numerical simulations, the annihilation scenario will show
the most cuspy profile at the GC direction.

\subsection{The WMAP haze}

The WMAP satellite has made all-sky survey in the microwave bands with
several frequencies in the range $\sim 10-100$ GHz. The Galactic
foreground dominates the radio emissions of these skymaps.
Such emissions are expected to be produced by standard interstellar
medium emission mechanisms, such as thermal dust, spinning dust,
ionized gas and synchrotron radiation. After excluding all known
contributions, however, Finkbeiner et al. reported an anomalous
excess of the microwave emission around the GC which is called
WMAP ``haze'' \cite{Finkbeiner:2003im}. In Ref. \cite{Dobler:2007wv},
the authors confirmed the existence of the haze by analyzing WMAP
three-year data and gave the possible implication of its origin.
It is also proposed that the haze might come from the DM annihilation
\cite{Hooper:2007kb}.

\begin{figure}[!htb]
\centering
\includegraphics[width=0.45\columnwidth]{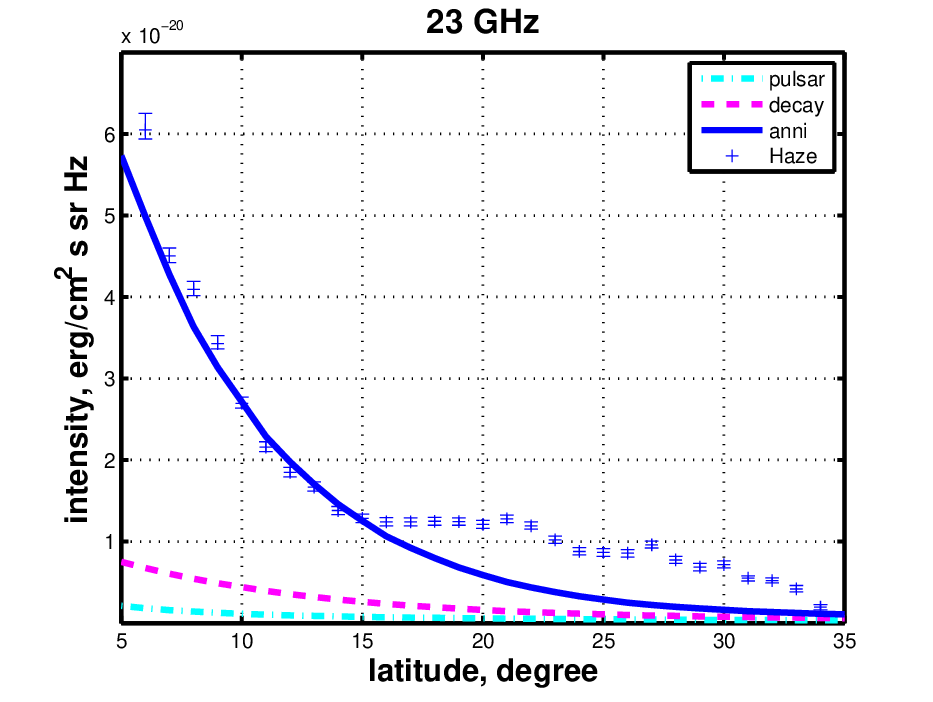}
\includegraphics[width=0.45\columnwidth]{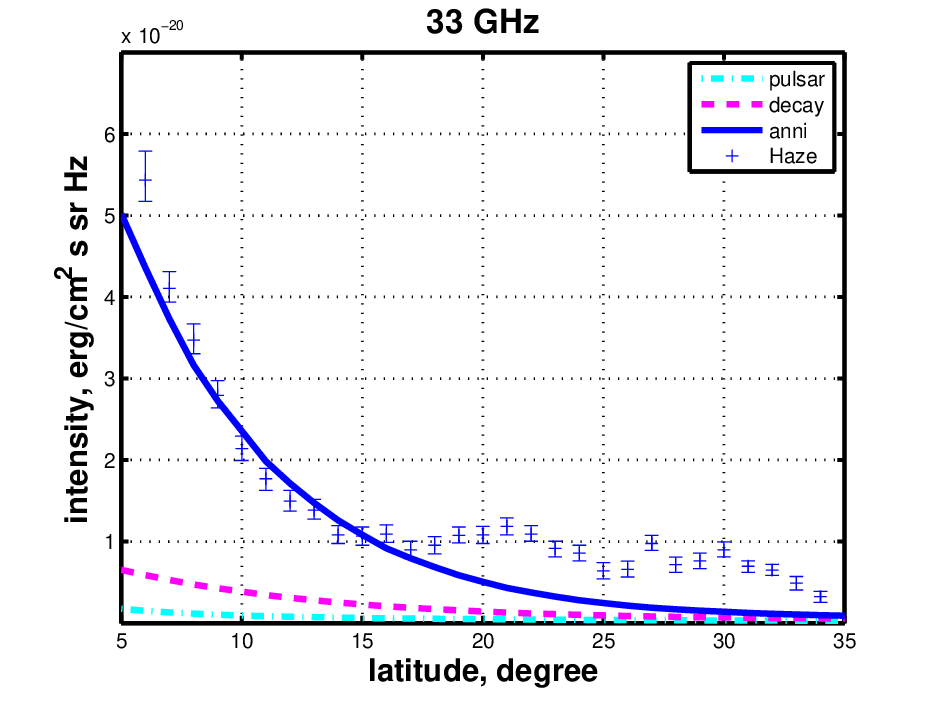}
\includegraphics[width=0.45\columnwidth]{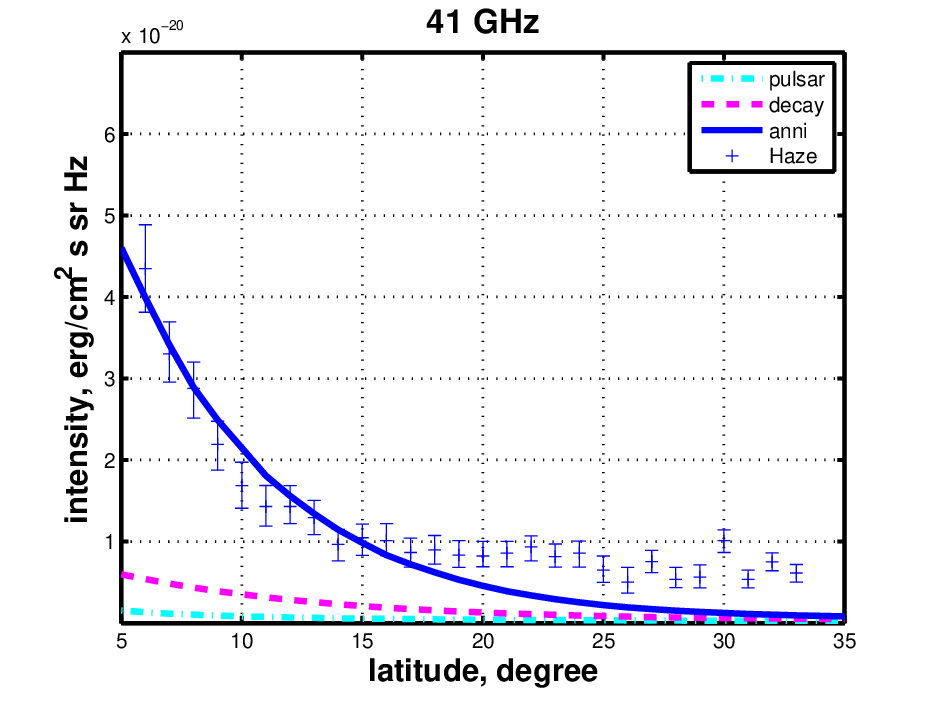}
\includegraphics[width=0.45\columnwidth]{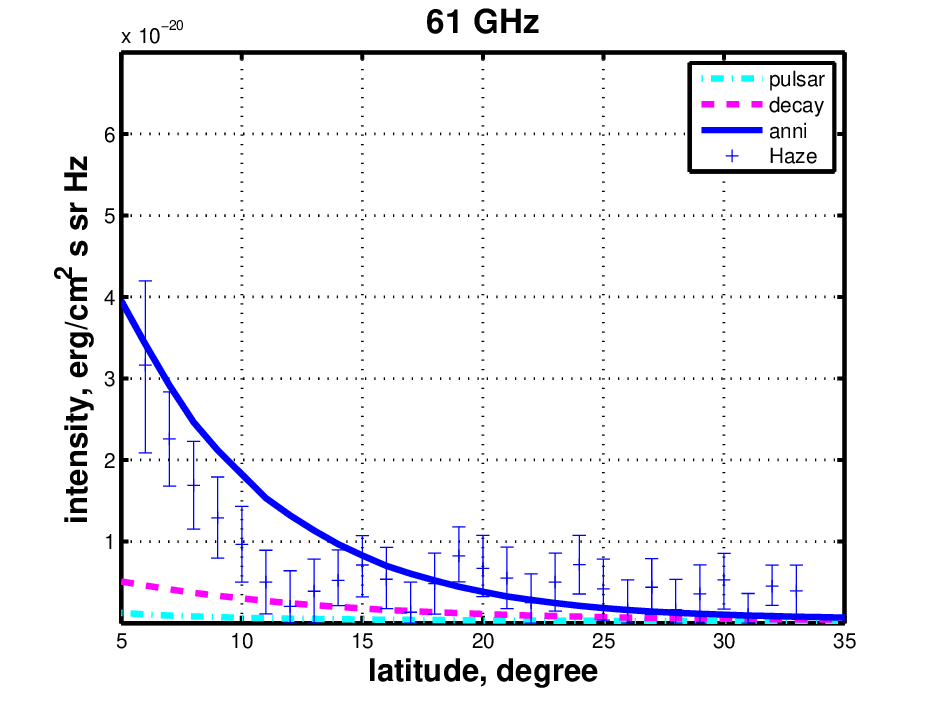}
\includegraphics[width=0.45\columnwidth]{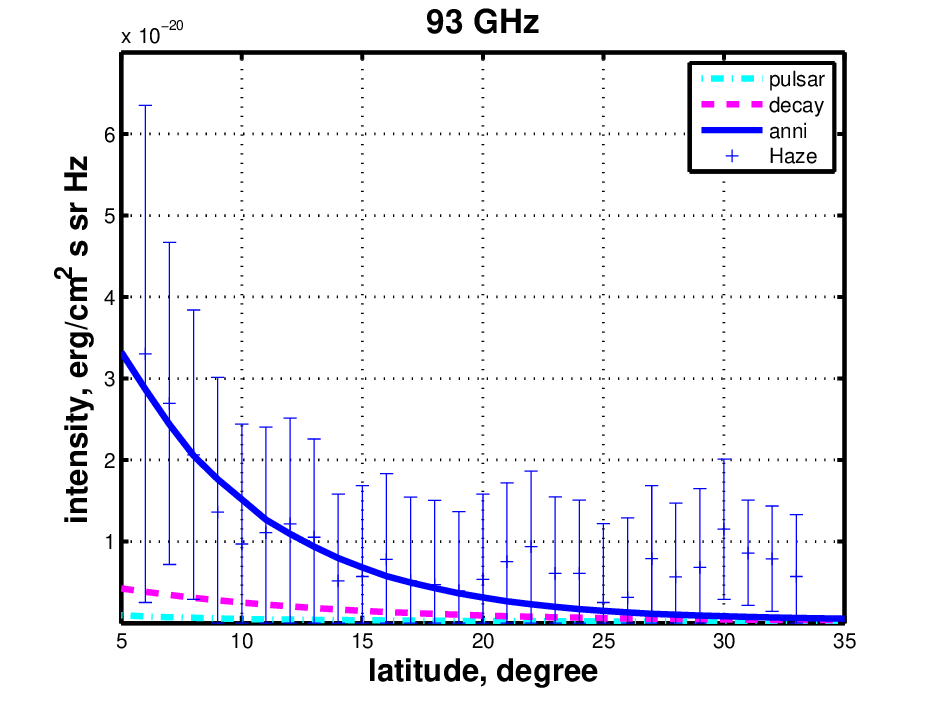}
\caption{
Residual synchrotron intensities after subtracting the background for
the three scenarios, compared with WMAP Haze data \cite{Dobler:2007wv} 
at five different frequencies. 
} \label{haze}
\end{figure}

In Figure \ref{haze} we compare our calculated synchrotron
profiles in the inner Galaxy from the three scenarios with the
WMAP data. From this figure we can see that annihilating DM model
may reproduce the WMAP haze data by taking proper density profile,
while the decaying DM and pulsar models can not produce the steep
rise toward the GC as shown in data.

\section{Diffuse $\gamma$-rays from IC scattering}

In this section we study the diffuse $\gamma$-ray emission in the
same three scenarios as discussed in the last two sections.
 Diffuse $\gamma$-rays can be produced through
IC scattering of energetic electrons and positrons with the
interstellar radiation field (ISRF). The ISRF consists of radio
emission of cosmic microwave background (CMB), infrared emission
from dust and optical emission from stars. A model to calculate
the ISRF distribution based on the realistic Galactic stellar and
dust distributions was built by Porter and Strong
\cite{Porter:2005qx} and this result has been included in the
GALPROP package. In this work we use GALPROP to calculate the
IC-induced diffuse $\gamma$-rays generated by the primary
electrons/positrons as well as the background diffuse
$\gamma$-rays generated by conventional CRs \cite{Strong:1998fr}.
Note that the final state gamma ray radiation of annihilating or
decaying DM by internal bremsstrahlung, which is significant near
the mass of DM, is not taken into account here. As shown in Ref.
\cite{Cholis:2008wq} the IC component will dominate the diffuse
$\gamma$-rays in the Fermi energy range, which is also of major
interest in this work.

\begin{figure}[!htbtb]
\centering
\includegraphics[width=0.45\columnwidth]{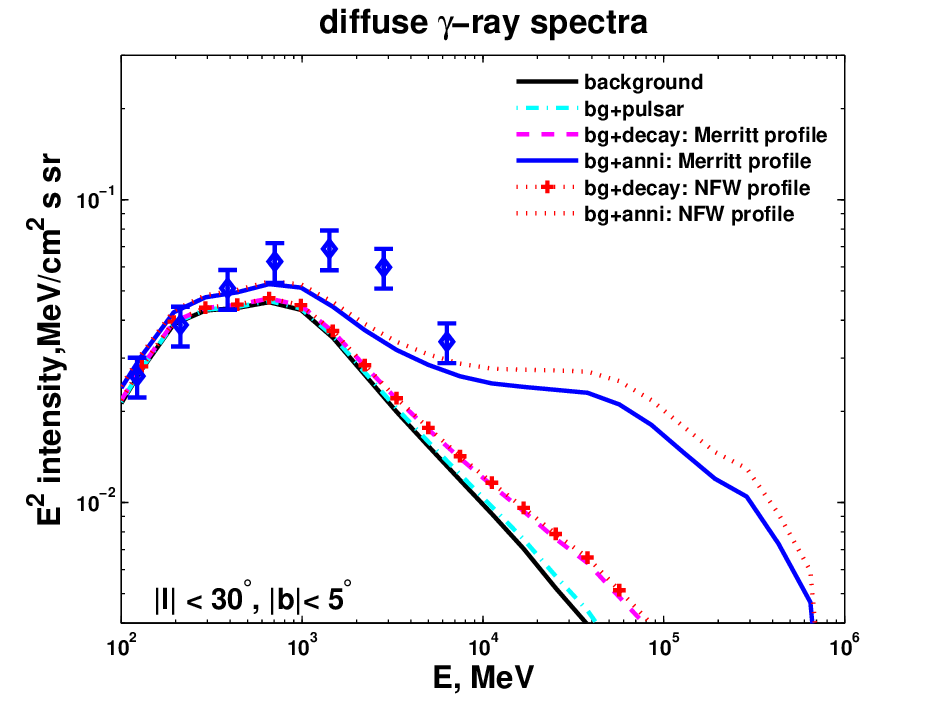}
\includegraphics[width=0.45\columnwidth]{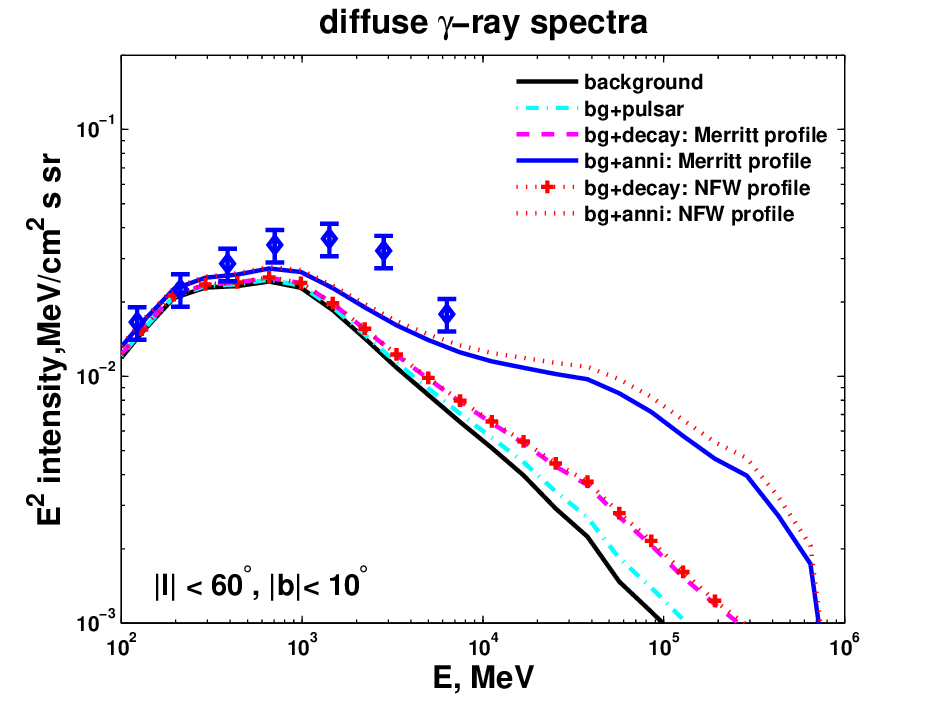}
\caption{The contributions to the diffuse $\gamma$-ray spectra in the
inner Galaxy of the three scenarios for two sky regions,
compared with the EGRET data \cite{Hunter:1997we}.
} \label{egret_A}
\end{figure}

In Figure \ref{egret_A}, we show the diffuse $\gamma$-ray spectra
for two sky regions, $|l| <30^\circ, |b| < 5^\circ$ and $|l|
<60^\circ, |b| < 10^\circ$ for the three scenarios,
corresponding to the region A \cite{Strong:2004de} and region H
\cite{Hunter:1997we} when discussing the EGRET data. The IC $\gamma$-rays
from DM annihilation are significantly larger than the other two
scenarios, especially for energies higher than several GeV. The
contributions from pulsars are very small and almost undetectable
if the uncertainties of the background are taken into account. The
signal of decaying DM model is moderate.
The results
from all the three scenarios do not violate the observations of
EGRET. 

\begin{figure}[!htb]
\centering
\includegraphics[width=0.45\columnwidth]{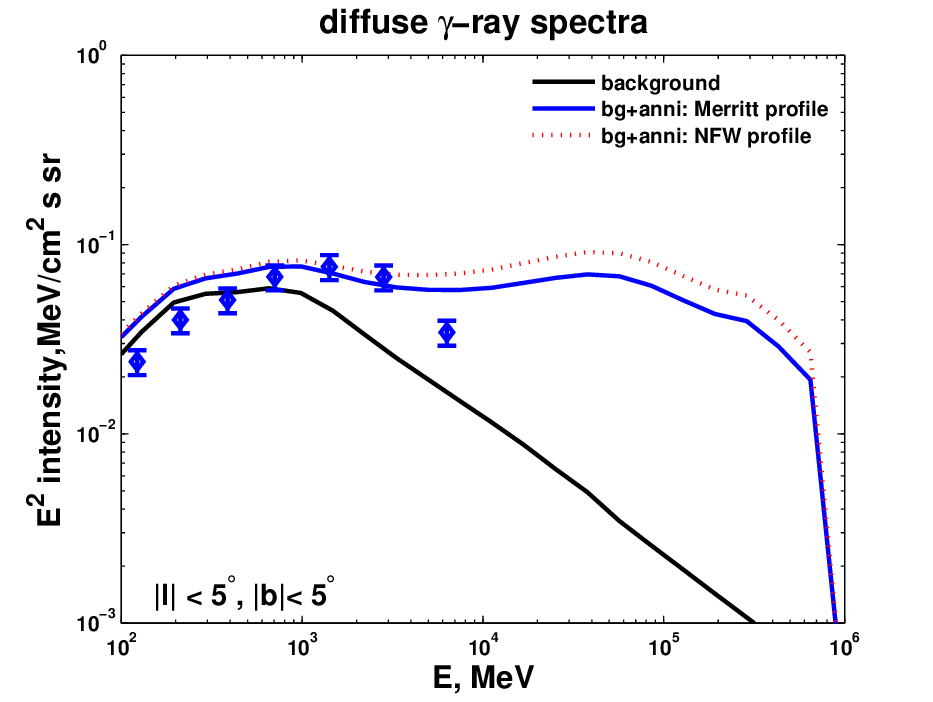}
\caption{Same as Figure \ref{egret_A} but for annihilation DM
within a smaller region $|l| <5^\circ, |b| < 5^\circ$.
NFW and Merritt profiles are adopted to show the difference.
} \label{egret_5}
\end{figure}

For comparison we also plot in Figure \ref{egret_A} the contributions
of annihilating DM with Navarro-Frenk-White (NFW) profile,
\begin{equation}
\label{nfwpf}
\rho(r)=\frac{\rho_{s}}{
\frac{r}{r_s}\left[ 1+\frac{r}{r_s} \right]^2},
\end{equation}
with $r_s=20$ kpc and the local DM density
$\rho_{\odot}=0.3$ GeV cm$^{-3}$
at $r=R_{\odot}\equiv8.5\,$kpc. 
Since NFW profile is steeper
in the halo center than Merritt profile, it will give harder $\gamma$-ray
spectra. In a smaller sky region, the differences between
these two kinds of DM distribution profiles are more obvious, as shown
in Figure \ref{egret_5}.
We can see that the IC $\gamma$-rays and synchrotron radiation can provide
a good test of the annihilating DM scenario and even give constraints on the
inner profile of DM distribution \cite{Bertone:2008xr,Bergstrom:2008ag}.

\begin{figure}[!htb]
\centering
\includegraphics[width=0.45\columnwidth]{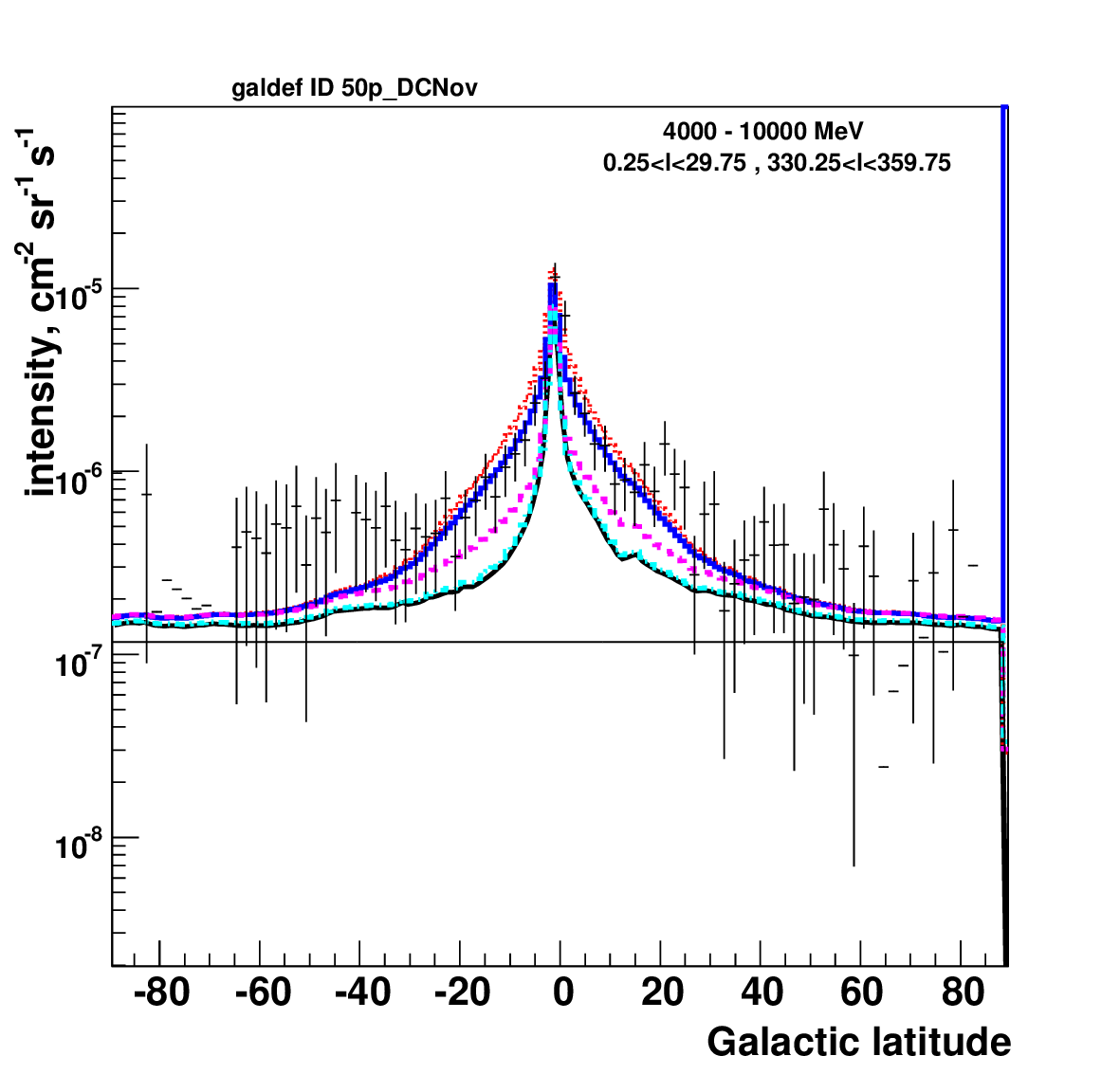}
\includegraphics[width=0.45\columnwidth]{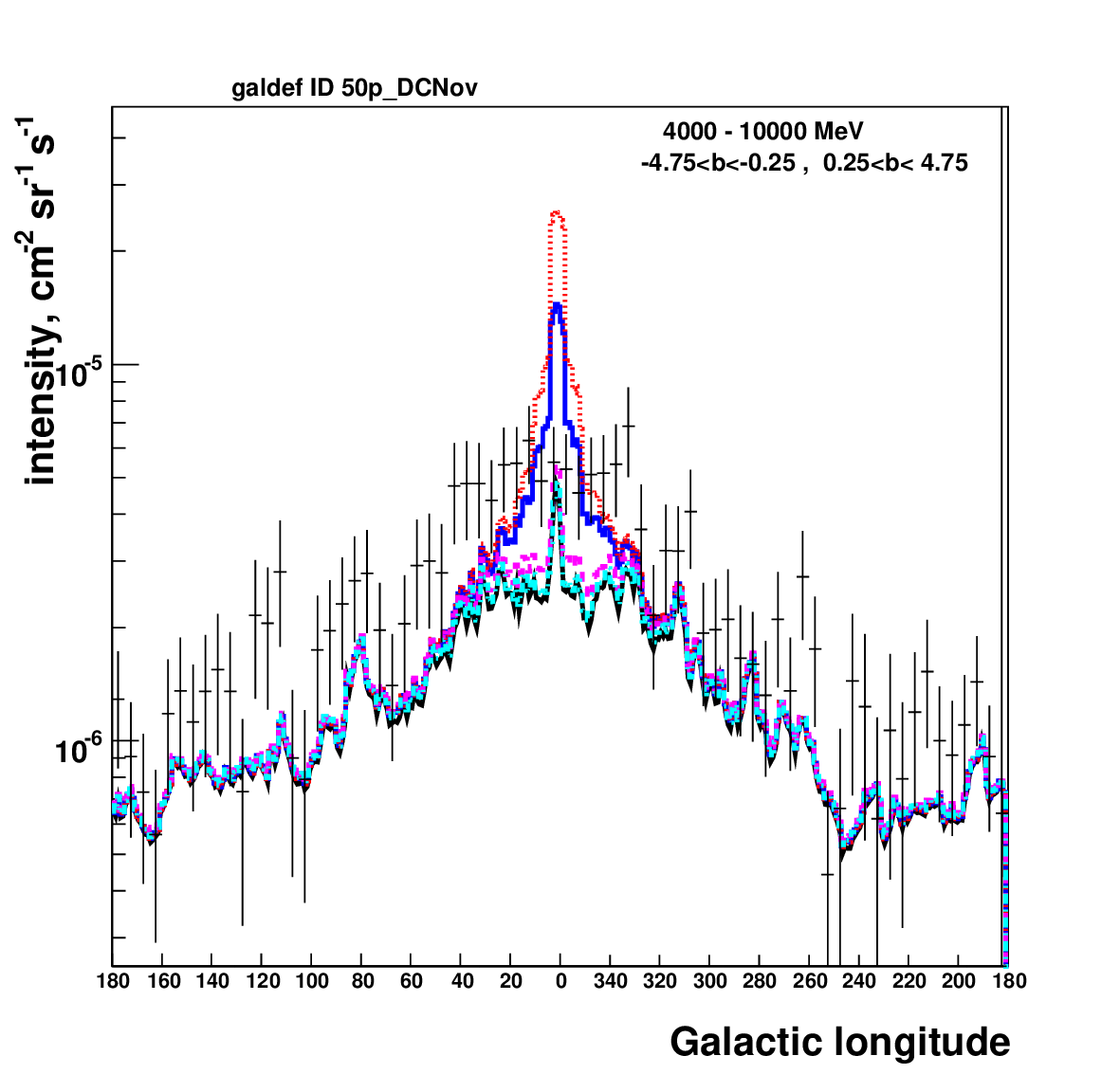}
\includegraphics[width=0.45\columnwidth]{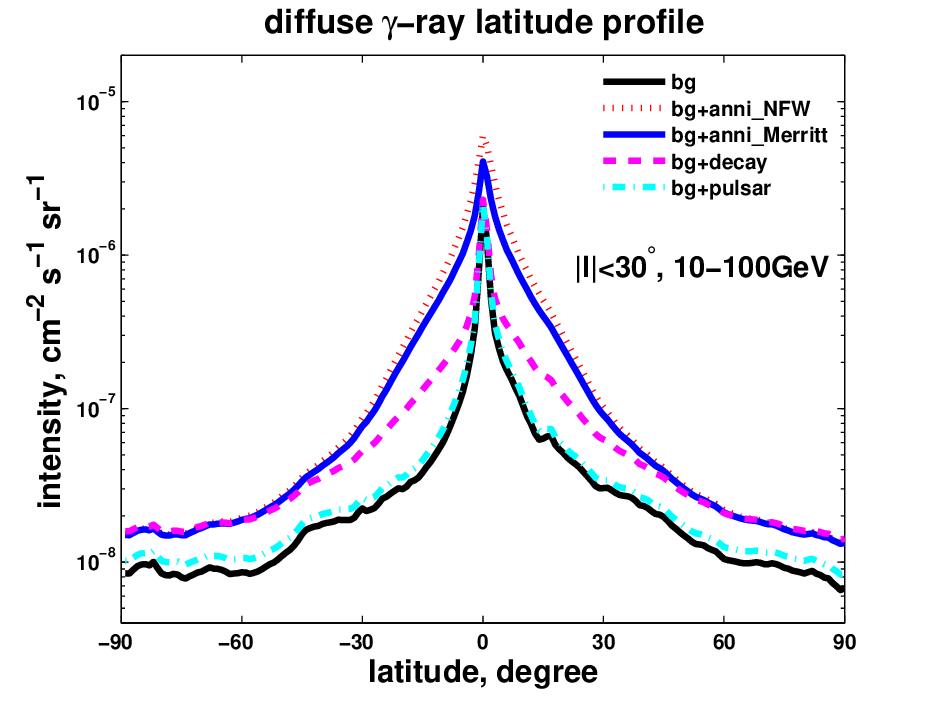}
\includegraphics[width=0.45\columnwidth]{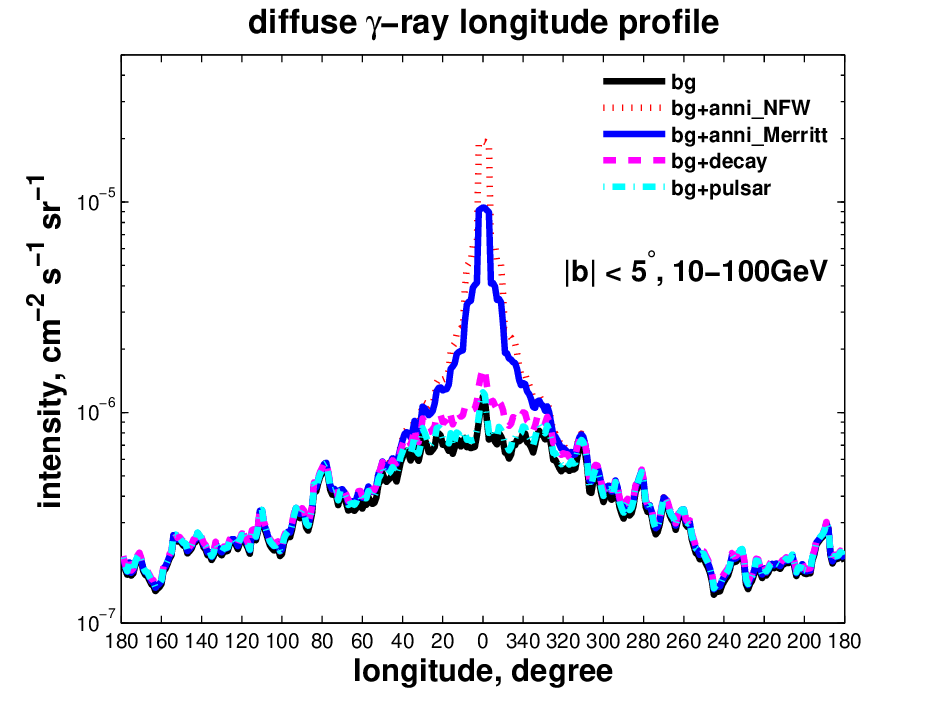}
\caption{The diffuse $\gamma-$ray latitude (average with $|l|<30^\circ$)
and longitude (average with $|b|<5^\circ$) profiles. {\it Top
two panels} are for energies $4-10$ GeV, and the {\it bottom two
panels} for $10-100$ GeV.
} \label{gamma_pf}
\end{figure}

Figure \ref{gamma_pf} gives the diffuse $\gamma$-ray longitude and
latitude profiles in the inner Galaxy, for two energy intervals,
$4-10$ GeV and $10-100$ GeV respectively. The data points in the
{\it top two panels} are from EGRET diffuse skymaps. It is shown that
the precise measurements of the $\gamma$-ray gradient near GC will
be also very helpful to discriminate different models, and possibly
reveal the particle properties of DM \cite{Jeltema:2008hf}.


\section{Discussion on the uncertainties of prediction}

In the last two sections, we study the prospects to discriminate
the three scenarios accounting for the PAMELA and ATIC data, i.e.
annihilating DM, decaying DM and Galactic pulsars, through the
synchrotron and IC radiation. For each scenario, a specific model
is chosen. In this section we will discuss the uncertainties of
the model configurations.

\subsection{Particle physics model of DM}

In this work we adopt the annihilation/decay modes of DM only to
leptons with equal branching ratios to $e^+e^-$, $\mu^+\mu^-$ and
$\tau^+\tau^-$ to explain the PAMELA and ATIC data. However, our
predictions of the synchrotron and IC radiation are not generally based
on this specific choice. Since no matter how a model is
constructed, the final electron and positron spectra have to be
adjusted to fit the PAMELA and ATIC data. It is the electron and
positron spectra that determine our prediction of the synchrotron
and IC radiation.

\begin{figure}[!htb]
\centering
\includegraphics[width=0.45\columnwidth]{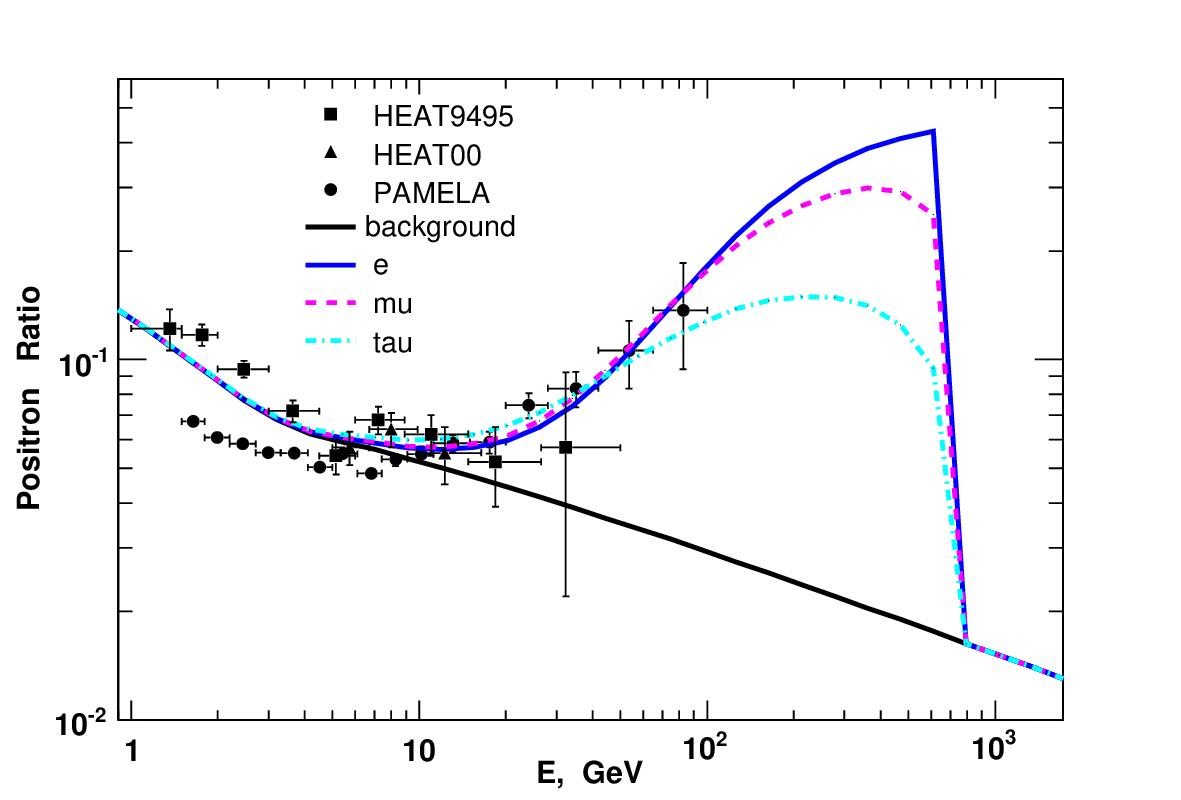}
\includegraphics[width=0.45\columnwidth]{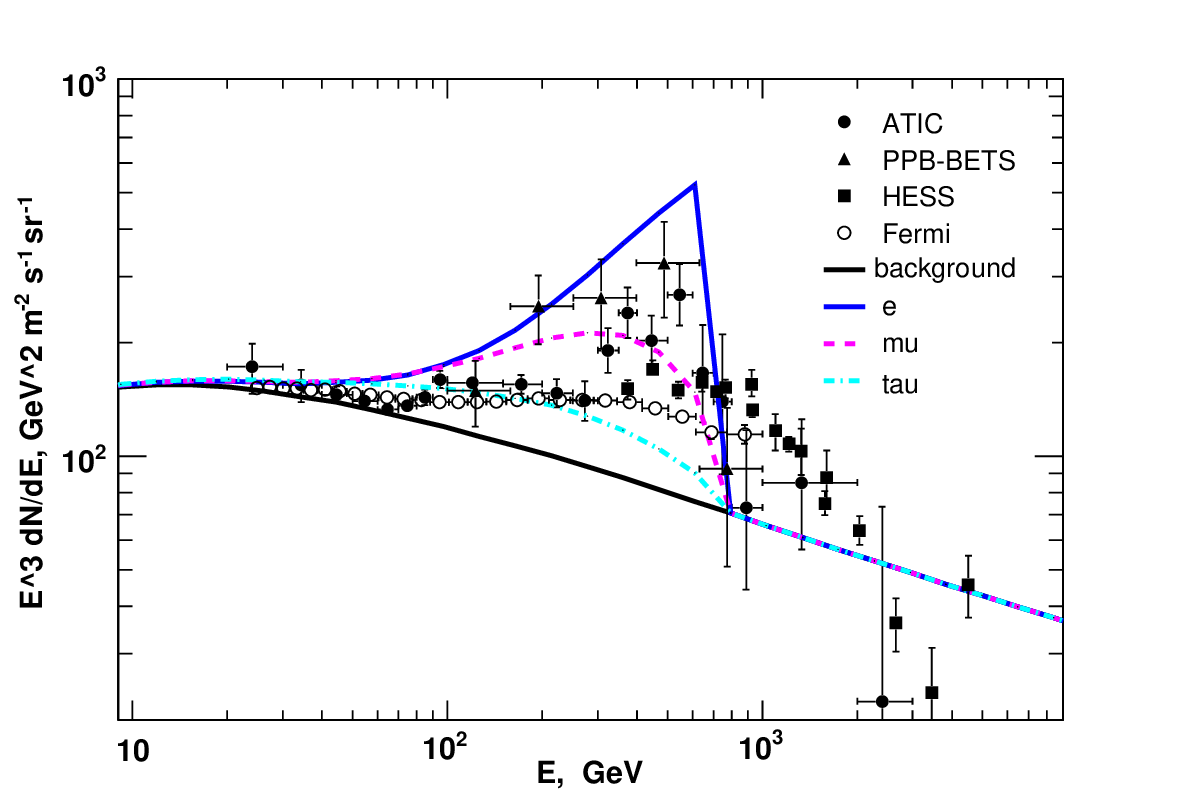}
\caption{The positron fractions and electron spectra in the case
of pure $e^+e^-$, $\mu^+\mu^-$ or $\tau^+\tau^-$ final state of
DM. } \label{e_mu_tau}
\end{figure}

\begin{figure} 
\centering
\includegraphics[width=0.45\columnwidth]{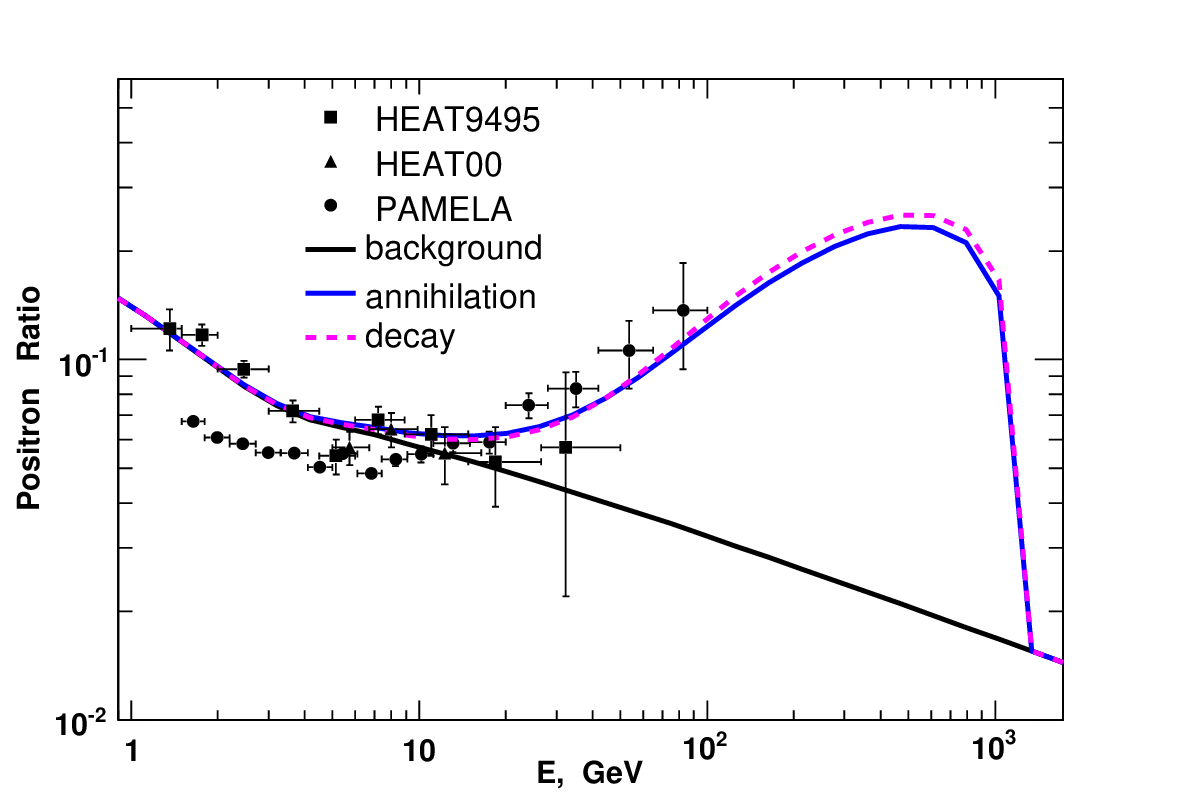}
\includegraphics[width=0.45\columnwidth]{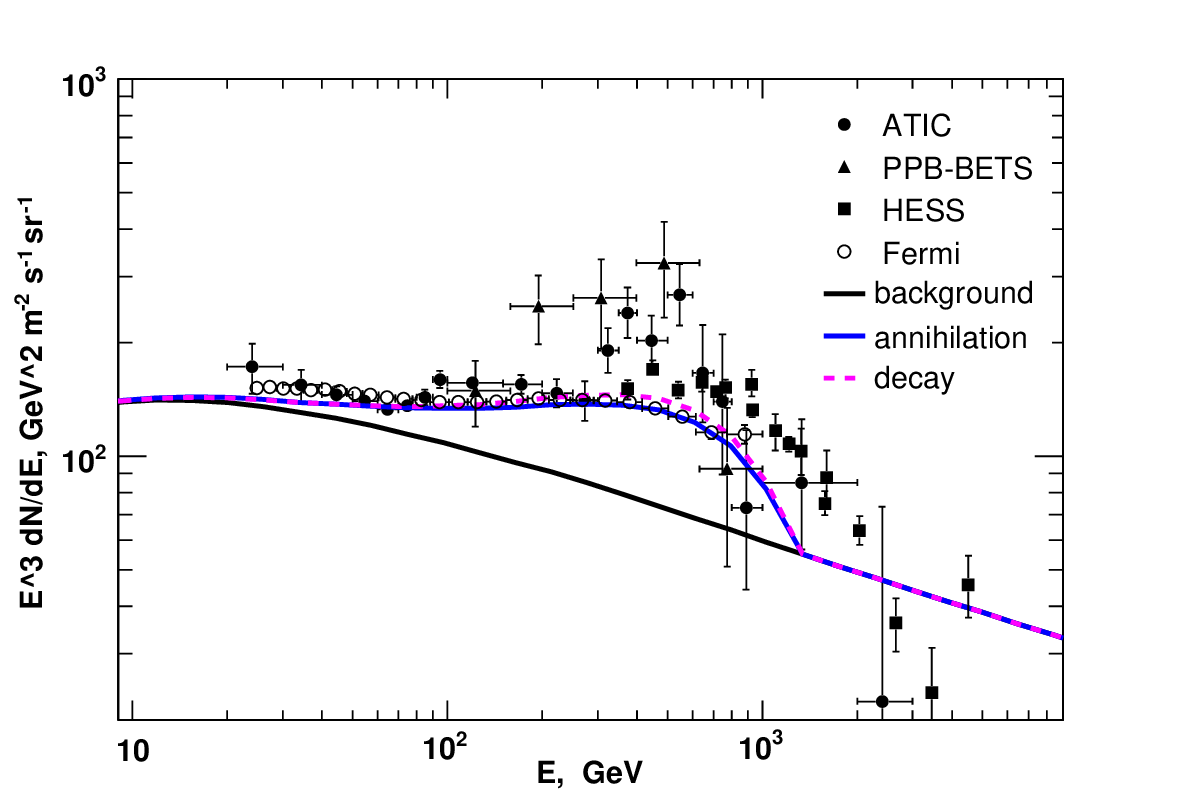}
\includegraphics[width=0.45\columnwidth]{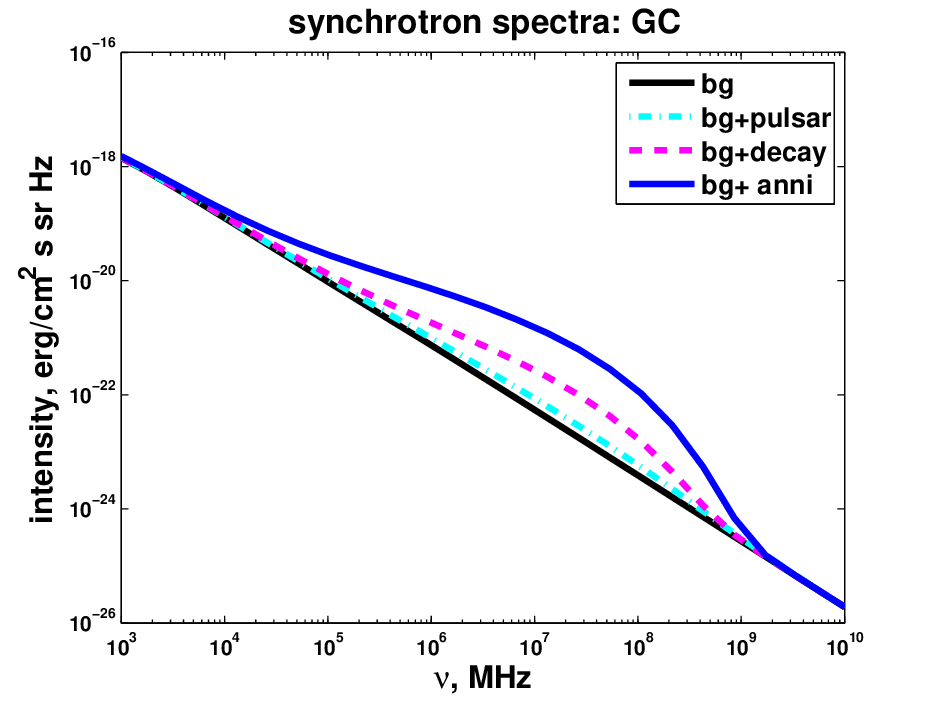}
\includegraphics[width=0.45\columnwidth]{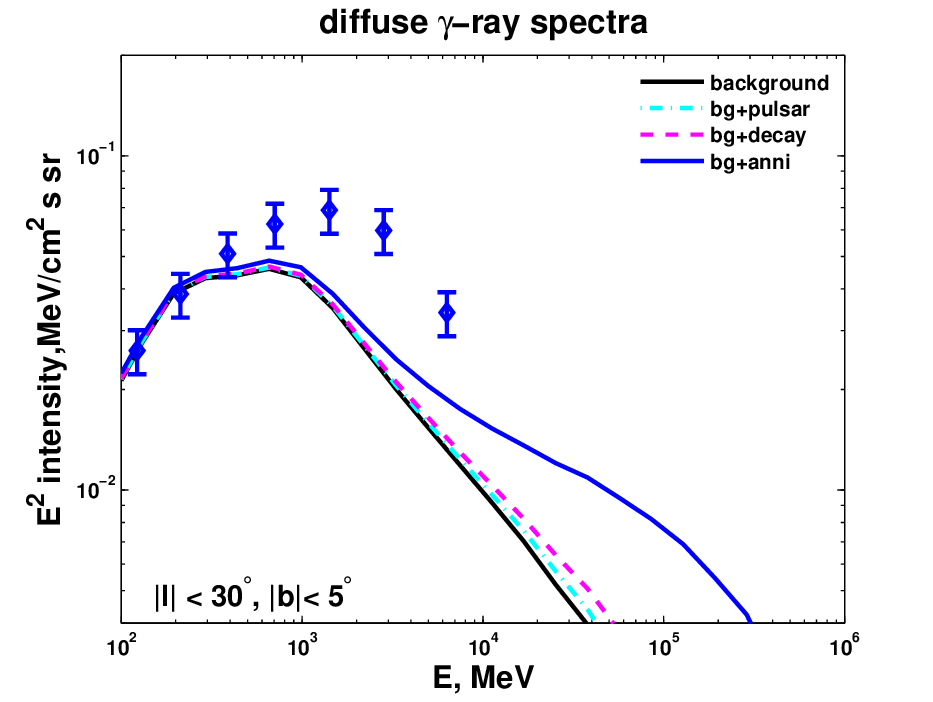}
 \caption{The predictions of the positron fraction (upper-left), total
(e$^++$e$^-$) spectrum (upper-right), synchrotron spectrum (bottom-left)
and diffuse $\gamma$-ray spectrum (bottom-right) for a 1.5 TeV DM annihilation model 
and a 3 TeV DM decay model, both of which 
with equal branching ratios into $\mu^+\mu^-$ and $\tau^+\tau^-$. The
pulsar scenario is the same as in Figures \ref{spectra} and
\ref{egret_A}.
} \label{1p5}
\end{figure}

Actually for a flavor blind DM the $1:1:1$ branching ratios to
$e^+e^-$, $\mu^+\mu^-$ and $\tau^+\tau^-$ is also the most natural
and typical choice. First the PAMELA antiproton data severely
constrained the gauge boson and quark final states
\cite{Adriani:2008zq}. As the first order approximation we assume
pure leptonic final states. Further, in Figure \ref{e_mu_tau} we
show that the PAMELA and ATIC data are not fitted very well if
only $e^+e^-$, $\mu^+\mu^-$ or $\tau^+\tau^-$ final state is
assumed. If we gauge the positron ratio to the PAMELA data, the
electron/positron spectrum will be too hard to account for the
ATIC data if only taking the $e^+e^-$ channel, while it is too
soft for the $\tau^+\tau^-$ channel. In addition, pure
$\mu^+\mu^-$ or $\tau^+\tau^-$ could not reproduce the sharp
falling at $\sim 600$ GeV at the ATIC spectrum. A combination of
the three channels is necessary to account for the data from the
two experiments simultaneously. The $e^+e^-$ channel is mainly to
produce the sharp falling at ATIC data while the $\tau^+\tau^-$
channel contributes to the PAMELA data at low energy.
According to the latest published data by Fermi
\cite{Collaboration:2009zk} and HESS \cite{Aharonian:2009ah}, the
bump feature of electron spectrum is not as prominent as that
observed by ATIC. 
In this case, the direct channel to $e^+e^-$ to
describe the fast drop of electron spectrum is not necessary. A DM
model with purely $\mu^+\mu^-$ as the final states has been
proposed to well fit the new data \cite{Bergstrom:2009fa}. While
from Fig. \ref{e_mu_tau} we can see that a pure $\tau^+\tau^-$
channel might also be able to fit the Fermi/H.E.S.S. data. In the
upper panels of Fig. \ref{1p5} we show the positron fractions and 
total (e$^++$e$^-$) spectra from DM annihilation and decay, of which the final
states have equal branching ratios into $\mu^+\mu^-$ and
$\tau^+\tau^-$ \footnote{A pure $\mu^+\mu^-$ or $\tau^+\tau^-$
channel can yield similar results, with the photon emission almost
unchanged.}, such as in the gauged $L_\mu-L_\tau$ model
\cite{Bi:2009uj}. In this calculation we adopt 
the same NFW DM density profile as in Eq.\ref{nfwpf}, 
$m_{\rm DM}=1.5$ TeV for annihilation and $m_{\rm DM}=3$ TeV for decay. 
The predicted synchrotron and diffuse $\gamma$-ray
spectra from the GC region are shown in the bottom panels of
Figure \ref{1p5}. We can see from this figure that the
annihilation DM model could still be discriminated. 

It should be noted that the current measurements of ATIC 
and Fermi are not fully consistent. Fermi result has high 
precision, however, both are dominated by systematic errors. 
The Fermi data seem to favor an astrophysical origin, such as 
from nearby pulsars. However, the ATIC data favor the DM 
scenario more. From our study we notice that the DM annihilation 
scenario seems always be distinguishable from the other two.


\subsection{Spatial distribution of DM}

The key point of the present work is that the three scenarios have
quite different spatial distributions, especially near the Galactic
center, although all the three scenarios are easily to account for
the local measurements by PAMELA and ATIC. It may be difficult to
discriminate the three scenarios by local observations near the
Solar system. However, the difference of the three scenarios near
the GC should be large if we can try to observe their products,
such as the synchrotron or IC radiation as we studied in this
work. Therefore, the most important uncertainties should come from
the spatial distribution of DM density.

In order to remove the uncertainties of the DM density profile we
have averaged the synchrotron and IC radiation  over a large
window around the Galactic center. This will smooth the signals
around the GC and diminish the uncertainties from DM density
profile efficiently as well as the distribution of pulsars.

Numerical simulations generally suggest cuspy DM profiles in the
center of the DM halo (e.g., \cite{Navarro:1996gj,Moore:1999gc,
Diemand:2005vz}). In this work we adopt the Merritt profile with
intermediate central density between the cuspy profiles like NFW
and the cored one like isothermal. This DM profile gives a good
description of the WMAP haze data, as shown in Figure \ref{haze}.

\begin{figure}[!htb]
\centering
\includegraphics[width=0.45\columnwidth]{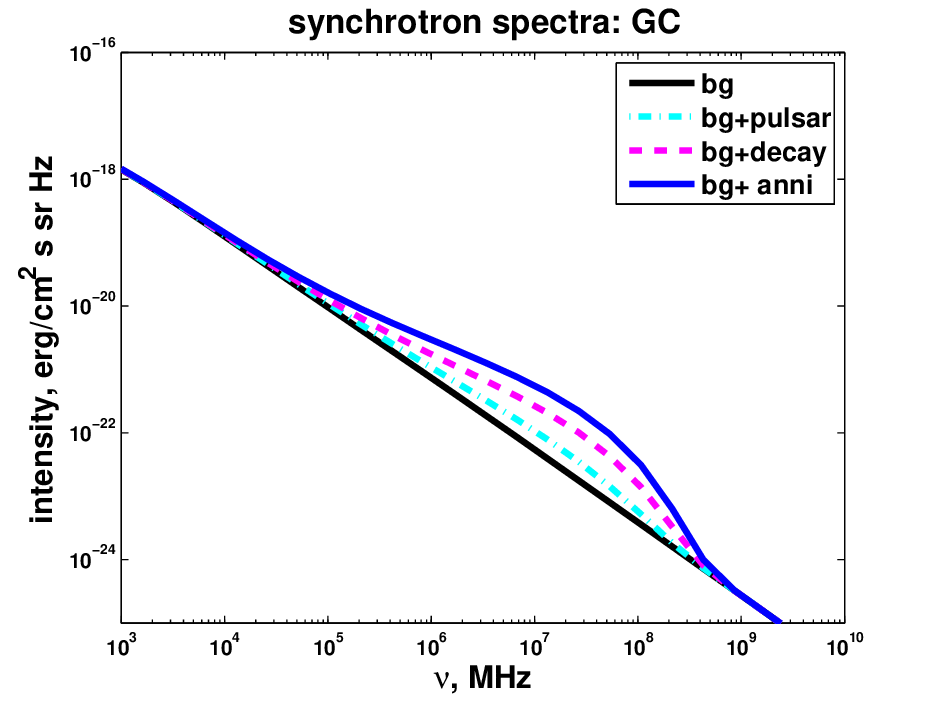}
\includegraphics[width=0.45\columnwidth]{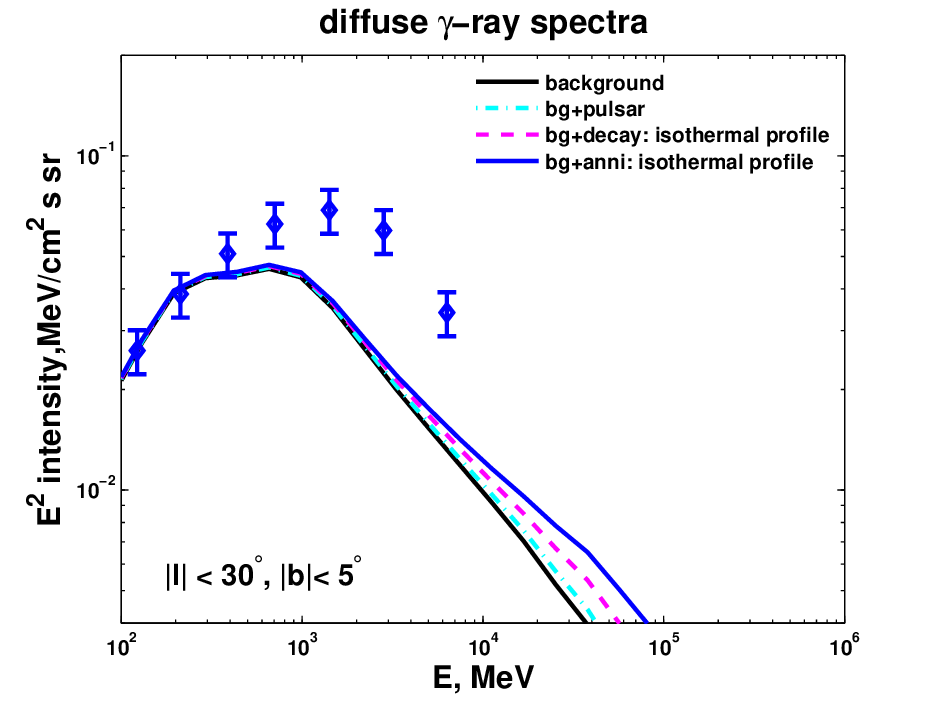}
\caption{The same as Figure 3 but for the isothermal DM profile.}
\label{isothermal}
\end{figure}

To show the uncertainty of DM profile here we consider the most
conservative case with the cored isothermal profile. The density
profile is given as
\begin{equation} \rho(r)=\frac{\rho_{0}}{ 1 + \left( \frac{r}{a}
\right)^2},
\end{equation}
with $a=5$ kpc and the local DM density $\rho_{\odot}=0.3$ GeV
cm$^{-3}$ at $r=R_{\odot}\equiv8.5\,$kpc. With the same local
density $\rho_{\odot}$, the cored isothermal profile is the least
cuspied distribution. Thus we anticipate it will produce the
weakest photon signals at the GC. We show in Figure
\ref{isothermal} the synchrotron spectra within a bin size of
$20^\circ\times20^\circ$ around the GC and the diffuse
$\gamma$-ray spectra around the GC for cored isothermal DM profile
compared with the pulsar model and the background. It is shown the
differences between these models indeed become smaller compared
with that given in Figure
\ref{spectra}.  
The results shown in
Figure \ref{isothermal} could be taken as the lower limit of our
prediction, if other related models and parameters are fixed.
Any other DM density profile will lead to a more clear
discrimination of these predictions. 
In this conservative case with cored isothermal profile, 
it might be not easy to discriminate the DM signals. 


\subsection{Propagation model}

Another source of uncertainties of our prediction comes from the
propagation models. Actually requiring the propagation model to
reproduce all the CR data gives strong constraints on the
propagation parameters. The secondary to primary ratio, such as
B/C, the unstable secondary to stable secondary, such as
$^{10}$Be/$^{9}$Be, and the local proton, electron data are used
to determine these parameters (see the discussion of the
propagation model in Ref \cite{Yin:2008bs}).
Under these requirement our results given in the previous sections
will not change much by adjusting the propagation parameters,
since the propagation models give similar descriptions to all
kinds of CRs. Once other species satisfy data the model should
also give similar electron and positron distributions at the same
time. We will explicitly show the expectation in the following.

In the previous sections we adopt the conventional model that all
the cosmic ray spectra are consistent with observations. We have
ignored the effect of reacceleration in Eq. (\ref{prop}) and kept
the effect of convection, which leads to better fit to the PAMELA
data at low energies \cite{Yin:2008bs}. We will show in the
following the propagation parameters will modify our results very
slightly.

\begin{figure}[!htb]
\centering
\includegraphics[width=0.45\columnwidth]{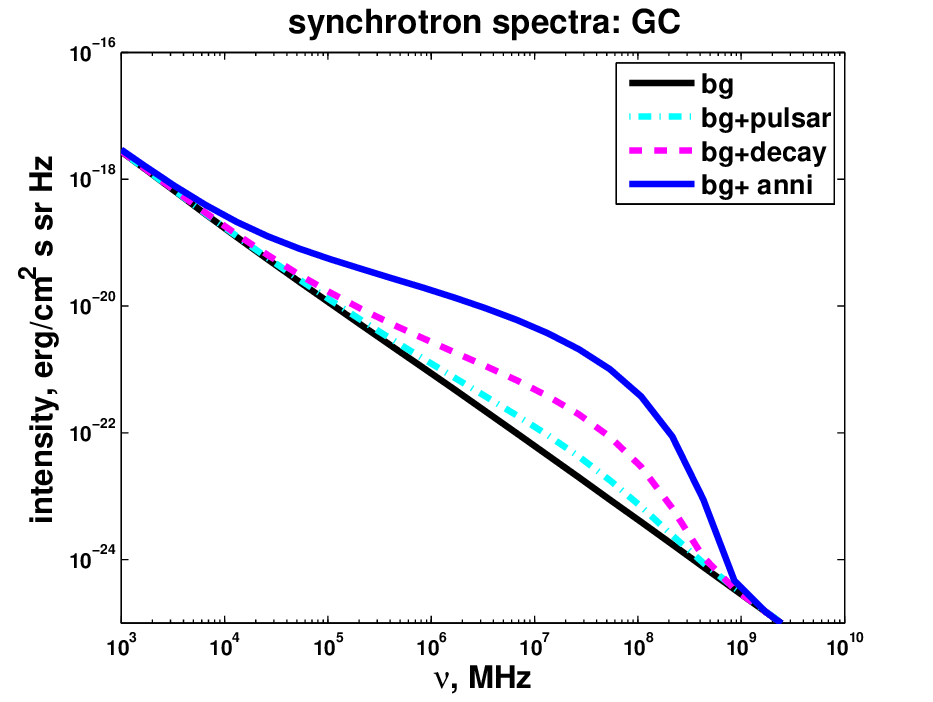}
\includegraphics[width=0.45\columnwidth]{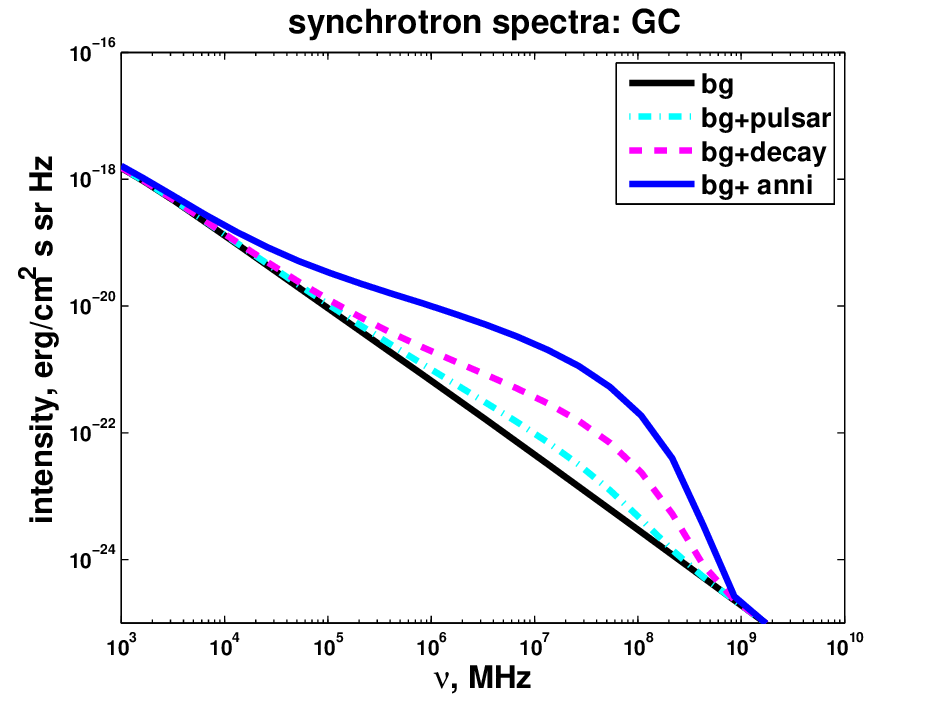}
\caption{The same as Figure 3 for {\it Left panel:} DR propagation
model and {\it Right panel:} DC propagation model but with 2kpc
diffuse halo height.} \label{DR}
\end{figure}

As an example we give the synchrotron spectra for diffusion +
reacceleration (DR) propagation model adopting the same
propagation parameters as in \cite{Yin:2008bs}, which, contrary to
the propagation model adopted in the present work, keeps the
effect of reacceleration while ignores the effect of convection.
In Figure \ref{DR} we plot the synchrotron spectra in the DR
propagation model for the three scenarios to account for the
PAMELA and ATIC data. We see that these results are almost the
same as the prediction in the DC model shown in the left panel of
Figure \ref{spectra}.

Another important uncertainty of propagation model comes from the
height of the diffusion region. We have studied a propagation
model with diffusion height 2kpc. The propagation parameters are
also adjusted to reproduce all the cosmic ray observations. In the
right panel of Fig. \ref{DR} we show the synchrotron predictions
in this 2kpc diffuse halo height propagation model. We see that
the change of propagation parameters generally affect our
predictions slightly.


\subsection{Other astrophysical uncertainties}

The synchrotron and IC radiation directly depend on the magnetic
field and ISRF respectively. The GALPROP model has adopted the
latest result of the magnetic field and ISRF, which are given
according to the most recent astronomical observations. To fully
figure out the uncertainties of these astronomical observations is
not easy. However, taking these uncertainties into account will
not change our conclusion in this work, since the change will
affect all the three scenarios simultaneously. For example the
strength of the magnetic field may have the uncertainty of a
factor $2$. To reduce the magnetic field twice will reduce all the
synchrotron radiation of the three scenarios, together with the
background, twice. This is equivalent to say that all the curves
in the synchrotron plots should shift downwards, while the
relative differences between these models are not changed.
Similarly change of ISRF will shift the IC radiation for all the
three scenarios at the same time. However, observations of radio
or diffuse $\gamma$-rays by EGRET and Fermi especially at high
latitude will constrain the magnetic field or ISRF that can not be
too small.

\section{Summary and conclusion}

To explain the recent measurements of the positron fraction by
PAMELA and electron spectrum by ATIC, Fermi and HESS, several
scenarios including the DM and pulsars are proposed. In this paper
we studied the perspective to discriminate these models using the
synchrotron and IC radiation generated by these
electrons/positrons. The point is that although various kinds of
models degenerate in the local environment\footnote{High energy
electrons lose energy fast and the detected electrons should
indeed be ``local'', e.g., within $\sim 1$ kpc
\cite{Maurin:2002uc}.} and give similar
electrons/positrons spectra, their different spatial extension might be 
revealed by the accompanied photon emission.

By properly adjusting the model parameters we first build three
benchmark models which can reproduce the measured
electron/positron data. 
We then studied the synchrotron and IC
radiation from these primary electrons and positrons. We find that
around the GC region, the synchrotron spectra for frequencies from
$10^4 \sim 10^9$ MHz and the IC $\gamma$-ray spectra from several
GeV to several hundred GeV are significantly different among the
three scenarios, especially between annihilating DM scenario and
others. For the directional profiles, the DM models show larger
gradient near the GC, especially in the longitude profiles, than
the pulsar model and the background. The photon emissions for
pulsar model are very small and similar with the background, and
will be the most difficult one to be detected. While for DM
models, especially the annihilating DM model, there are strong
observable signals. Finally we discuss the possible uncertainties
of these conclusions. The major uncertainty comes from the DM
inner profile. For the most conservative case with cored
isothermal DM profile, the signals of DM scenarios become weaker
and might be hard to be discriminated. However, the isothermal
profile is generally disfavored by numerical simulations. Other
uncertainties, such as the DM annihilation or decay channels, the
propagation parameters and other astrophysical inputs, seems not
to change the relative differences among the signals of the three
scenarios significantly.

It should be pointed out that in this work the boost factor for
the annihilating DM scenario is assumed to be universal, such as
the Sommerfeld effect. It is also possible that the boost effect
is spatially dependent, e.g., from the DM subhalos
\cite{Yuan:2006ju,Bi:2007ab,Bi:2006vk}. If so, the enhancement
near the GC region will not be as important as at large radius in
the halo. Actually, the calculation based on the N-body simulation
indicates that the boost factor from DM substructure for
anti-matter particles from DM annihilation in the Galaxy is
generally negligible \cite{Lavalle:1900wn,Lavalle:2008zb}.
A recent calculation on the DM clumpiness boost
factor taking into account the Sommerfeld effect still finds no
remarkable enhancement effect, except for fine tuning to the
strongly resonant case \cite{Yuan:2009bb}. Therefore the conclusions
in this work are generally held.

\acknowledgments

We thank Prof. Fang-Jun Lu for helpful discussion on pulsars and
PWNe. This work is supported in part by the Natural Sciences
Foundation of China (Nos. 10773011, 10775001, 10635030), by the
trans-century fund of Chinese Ministry of Education, and by the
Chinese Academy of Sciences under the grant No. KJCX3-SYW-N2.

\end{document}